\documentclass[twocolumn,trackchanges]{aastex631}

\newcommand{\kms}{\hbox{km\,s$^{-1}$}} 
\newcommand{\logg}{\hbox{log\,$\it g$}}

\newcommand{\teff}{\hbox{$T_{\rm eff}$}}

\usepackage{lmodern}
\usepackage[T1]{fontenc}
\usepackage{textcomp}
\usepackage{graphicx}
\usepackage{subfigure}
\usepackage{orcidlink}

\shorttitle{An Empirical Sample of M stars}
\shortauthors{Du et al.}

\begin{document}
\title{An Empirical Sample of Spectra of M-type Stars
with Homogeneous Atmospheric-Parameter Labels}

\correspondingauthor{ A-Li Luo, Song Wang}
\email{ lal@nao.cas.cn, songw@bao.ac.cn}
\author{Bing Du,$^{\orcidlink{0000-0001-6820-6441}}$}
\affiliation{National Astronomical Observatories, Chinese Academy of Sciences, Beijing 100101, China}
\author{A-Li Luo$^{\orcidlink{0000-0001-7865-2648}}$}
\affiliation{National Astronomical Observatories, Chinese Academy of Sciences, Beijing 100101, China}
\affiliation{School of Astronomy and Space Science, University of Chinese Academy of Sciences, Beijing 100049, China}
\affiliation{University of Chinese Academy of Sciences, Nanjing, Jiangsu, 211135, China}
\author{Song Wang$^{\orcidlink{0000-0003-3116-5038}}$}
\affiliation{National Astronomical Observatories, Chinese Academy of Sciences, Beijing 100101, China}
\affiliation{Institute for Frontiers in Astronomy and Astrophysics, Beijing Normal University, Beijing 102206, China}
\author{Yinbi Li$^{\orcidlink{0000-0001-7607-2666}}$}
\affiliation{National Astronomical Observatories, Chinese Academy of Sciences, Beijing 100101, China}
\author{Cai-Xia, Qu$^{\orcidlink{0000-0002-5460-2205}}$}
\affiliation{National Astronomical Observatories, Chinese Academy of Sciences, Beijing 100101, China}
\affiliation{School of Astronomy and Space Science, University of Chinese Academy of Sciences, Beijing 100049, China}
\author{Xiao Kong$^{\orcidlink{0000-0001-8011-8401}}$}
\affiliation{National Astronomical Observatories, Chinese Academy of Sciences, Beijing 100101, China}
\author{Yan-xin Guo$^{\orcidlink{0000-0002-7640-5368}}$}
\affiliation{National Astronomical Observatories, Chinese Academy of Sciences, Beijing 100101, China}
\author{Yi-han Song$^{\orcidlink{0000-0001-7255-5003}}$} 
\affiliation{National Astronomical Observatories, Chinese Academy of Sciences, Beijing 100101, China}
\author{Fang Zuo$^{\orcidlink{0000-0002-9081-8951}}$}
\affiliation{National Astronomical Observatories, Chinese Academy of Sciences, Beijing 100101, China}

\begin{abstract}
The discrepancies between theoretical and observed spectra, and the systematic differences between various spectroscopic parameter estimates,  complicate the determination of atmospheric parameters of M-type stars. In this work, we present an empirical sample of 5105 M-type star spectra  with  homogeneous atmospheric parameter labels through stellar-label transfer and sample cleaning. We addressed systematic discrepancies in spectroscopic parameter estimates by adopting recent results for Gaia EDR3 stars as a reference standard. Then, we used a density-based spatial clustering of applications with noise  to remove unreliable samples in each subgrid of parameters. To confirm the reliability of the stellar labels, a 5-layer neural network was utilized, randomly partitioning the samples into training and testing sets. The standard deviations between the predicted and actual values in the testing set are 14 K for \teff, 0.06 dex for \logg, and 0.05 dex for [M/H], respectively. In addition, we conducted an internal cross-validation to enhance validation and obtained precisions of 11 K, 0.05 dex, and 0.05 dex for \teff, \logg, and [M/H], respectively.  A grid of 1365 high Signal-to-Noise ratio (S/N) spectra and their labels,  selected from the empirical sample,  was utilized in the stellar parameter pipeline for M-Type stars (LASPM) of the  Large Sky Area Multi-Object Fiber Spectroscopic Telescope (LAMOST), producing an almost seamless Kiel distribution diagram for LAMOST \textbf{DR10 and DR11 data}. The atmospheric parameters for M-type stars from LAMOST DR11  show  improved precision compared to  the data from DR9, with improvements (for spectra with S/N higher than 10) from 118 to 67 K in \teff,  0.2 to 0.07 dex in \logg, and 0.29 to 0.14 dex in [M/H].

\end{abstract}



\section{Introduction} \label{sec:intro}
The atmospheric parameters of M-type stars offer valuable information on revealing the formation history of the Galaxy. The M dwarf stars, which dominate the faint magnitudes of the Galaxy \citep{1997ApJ...482..420L,2010AJ....139.2679B}, are not only important for determining the initial mass function \citep{2008AJ....136.1778C,2024ApJS..271...55K},  but also good tracers of the chemical and dynamical history of the Milky Way because of their exceptionally long lifetime. The M Giants with  high luminosity are good tracers for revealing the accretion and merger events in the Galaxy by discovering and identifying substructures in the Galactic outer disk and remnants of stellar streams in the halo \citep{2009ssc..book.....G,2014AJ....147...76B,2023ApJS..266....4L}. 

The stellar parameters of M-type stars such as effective temperature (\teff), surface gravity (log$g$), and metallicity ([M/H]) are model dependent. Impressive progress has been made in the study of atmospheric models and molecular absorption of late-type stars in the three decades. For example, the PHOENIX BT-Settl model, using revised solar abundances and updated atomic and molecular line opacities \citep{2012EAS....57....3A,2013MSAIS..24..128A}, can reproduce the observed spectra very well for M dwarfs \citep{2013A&A...556A..15R}. Consequently, the BT-Settl model grids were utilized by \cite{2021RAA....21..202D} to establish the stellar parameter pipeline for M-type stars (LASPM) of the Large Sky Area Multi-Object Fiber Spectroscopic Telescope (LAMOST; \citealt{2012RAA....12.1197C,2012RAA....12..735D}). However, the BT-Settl model has a poor fit to the observed spectra for M giants, making the LASPM parameters of M giants unreliable \citep{2021RAA....21..202D, 2023RAA....23e5008Q}. Compared with the BT-Settl model, the MARCS model has a good fit with the observed spectra  for M giants ranging from M1 to M6III, but it has an excess blue flux for M dwarfs \citep{2008A&A...486..951G,2008PhST..133a4003P}. 

Although substantial progress has been made with the models, they remain imperfect. For example,  \citet{2013A&A...556A..15R}  compared the observed colors to the BT-Settl colors,  and discovered offsets between the observed and theoretical colors in certain cases, along with a general dispersion.   \cite{2020A&A...640A..87M} derived atmospheric parameters for red giant branch stars both from the photometry and from excitation and ionisation balances. A discrepancy was found between the effective temperatures derived from the spectra and those derived from the photometic data,  and this discrepancy increases as the metallicity decreases. \cite{2013ApJ...767....3D} determined the effective temperatures of red supergiants by fitting MARCS model atmospheres to various parts of their optical and near-infrared spectra, deriving the effective temperatures for each star from (1) the TiO bands, (2) clear continuum regions of the spectral energy distributions (SEDs), and (3) the integrated fluxes. \cite{2013ApJ...767....3D} found that the temperatures obtained from fits to the TiO bands are systematically lower by several hundred kelvin compared to the other two methods and proposed that the \teff \ from optical and near-infrared SEDs are more representative. Precisely determining the atmospheric parameters of M-type stars remains challenging, especially with respect to metallicity. It is still difficult to obtain a consistent metallicity of M stars in the optical region from different models \citep{2016A&A...587A..19P}.

Because of the discrepancies between theoretical and observed spectra, there are systematic differences between  different spectroscopic parameter estimates. The LASPM overestimates \logg \ by 0.63 dex and underestimates [M/H] by 0.25 dex when compared to the catalog of the Sloan Digital Sky Survey Apache Point Observatory Galactic Evolution Experiment (SDSS/APOGEE) \citep{2021RAA....21..202D}. Maintaining the precision shown in \citep{2021RAA....21..202D},  the revised LASPM applied in the Data Release 9 (DR9), continues to overestimate \logg \ by 0.27 dex relative to APOGEE\footnote{\url{https://www.lamost.org/dr9/v2.0/catalog}}.  \cite{2022ApJS..260...45D} determined the parameters of M-type stars of LAMOST DR8 by applying the MILES interpolator to the ULySS package.  Systematic  differences still exist between the results of \cite{2022ApJS..260...45D} and those of the APOGEE Stellar Parameter and Chemical Abundances Pipeline (ASPCAP). The causes of systematic differences may be attributed to the imperfections of the theoretical models.

Thanks to efforts to decode the stellar parameters from high-resolution spectra \citep{2014A&A...564A..90R,2016A&A...587A..19P, 2017ApJ...851...26V,2018A&A...620A.180R}, the ASPCAP of SDSS Data Release 16 (DR16) determined effective temperatures down to 3000 K by using  new atmospheric grids \citep{2020AJ....160..120J}.  The parameters of late-type stars from high-resolution spectra enables the data-driven methods like the Cannon \citep{2015ApJ...808...16N},  to measure the stellar parameters of M-type stars from either low-resolution spectra or multi-band photometric data by transferring stellar labels. For example, \cite{2021ApJS..253...45L} determined $\sim$ 300,000  spectroscopic stellar parameters (\teff \ and [M/H]) of  M dwarfs by training a Stellar LAbel Machine (SLAM) model using the LAMOST spectra with APOGEE DR16 stellar labels. Employing the SLAM model, \cite{2023RAA....23e5008Q} determined the stellar parameters (\teff, \logg, [M/H], $\alpha$/M]) for over 43,000 M giants by utilizing the LAMOST spectra and APOGEE DR17 stellar labels. Building on the previous research, \cite{2024ApJS..270...32Q} obtained atmospheric parameters for 1,806,921 cool dwarfs using Gaia DR3, employing machine learning algorithms trained on multiband photometry and stellar labels from APOGEE DR16, along with  catalogs from \cite{2021ApJS..253...45L} and \cite{2022ApJS..260...45D}.

In addition to the spectroscopic parameters, photo-astrometric  parameters  were produced for Gaia EDR3 stars brighter than G =18.5  by \citet{2019A&A...628A..94A,2022A&A...658A..91A} using the StarHorse code, an isochrone-ﬁtting code that utilized  Gaia data combined with the photometric catalogs of Pan-STARRS1, SkyMapper, 2MASS, and AllWISE.  Thanks to the higher precision of the Gaia EDR3 and the new stellar-density priors of StarHorse, the precision over previous estimates was substantially improved with a typical precision of 3\% (15\%) in distance and 140 K (180 K) in \teff \ at magnitude $G = $ 14 (17) \citep{2022A&A...658A..91A}.  Although the photo-astrometric parameters are not as accurate as the spectroscopic  parameters, the majority of them are consistent with the stellar positions in both the color-magnitude diagram and the isochrones. Consequently,  the photo-astrometric parameters provided by StarHorse   can serve as a reference standard for adjusting the systematic discrepancies found between various collections of spectroscopic parameters.

By gathering parameter estimates  from LAMOST late-type stars and using the StarHorse catalog as a standard, it is possible to filter out an empirical sample with  homogeneous atmospheric parameter labels. The empirical sample can serve as an empirical library for the determination of stellar parameters and also provides data annotations for machine learning algorithms driven by data. Using this empirical sample as reference stars, the LASPM can to some extent circumvent the issue of mismatches between theoretical and observed spectra.

In this work, we present an empirical sample of  about 5000  M-type star spectra,  with  homogeneous stellar labels through stellar label transfer and sample cleaning. We address systematic discrepancies  between spectroscopic parameter estimates  employing  recent results for Gaia EDR3 stars as the reference standard.  Given the dense forest of spectral features of M-type stars in the optical band, we select clean samples in each parameter bin rather than co-adding spectra as outlined in \cite{2019ApJS..240...10D}. We remove  the unreliable samples in each subgrid of parameters.

A grid of 1365 high Signal-to-Noise ratio (S/N) spectra and their labels, chosen from  the  empirical sample,   is employed by LASPM for the determination of stellar parameters of M-type stars. The key algorithm of LASPM aims to find the best-agreement template from the reference library for a given observed spectrum, minimizing $\chi^2$ by linearly combining the ﬁve best-agreement templates.    The updated LASPM,  utilizing the new sample and its labels as a reference,  provides data products for the LAMOST data releases DR10 and DR11. This paper is organized as follows. In Section \ref{sec: sample},  we detail sample selection and cleaning  and demonstrate the verification of the stellar labels for our samples.  In Section \ref{sec:results}, we present the results of this empirical sample, including the stellar parameter coverage and the spectra.  In Section \ref{sec:application}, we use the new sample and its labels as a reference to reevaluate the parameters of the full sample of LAMOST M-type spectra and discuss the results.  Finally, the main conclusions are summarized in Section \ref{sec: summary}.

\section{Sample selection and cleaning} \label{sec: sample}

 The LAMOST low-resolution survey has gathered over 10 million spectra (R $\sim$ 1800, 3800--9000 \AA). The spectra were analyzed by the LAMOST 1D pipeline to recognize their spectral classes and determine the radial velocity (RV) for stars \citep{2015RAA....15.1095L}. For LAMOST DR11, the LAMOST 1D pipeline recognized more than 1 million M-type stars with spectral types of M1--M9 and roughly luminosity classes of M giants (gM) and M dwarfs (dM). Out of these, LASPM analyzed 877,570 M-type spectra with a spectral S/N $\geq$ 5.0 in the $i$-band (corresponding to 667,483 stars) to determine their stellar parameters. In contrast to the parameters released in DR7 and DR8 \citep{2021RAA....21..202D}, the systematic discrepancy in \logg \ for dwarfs was reduced after DR8 because we fine-tuned the resolution of the synthetic BT-Settl spectra used by the revised LASPM.

\begin{figure}[ht!]
	\centering
	\includegraphics[width=1.0\linewidth]{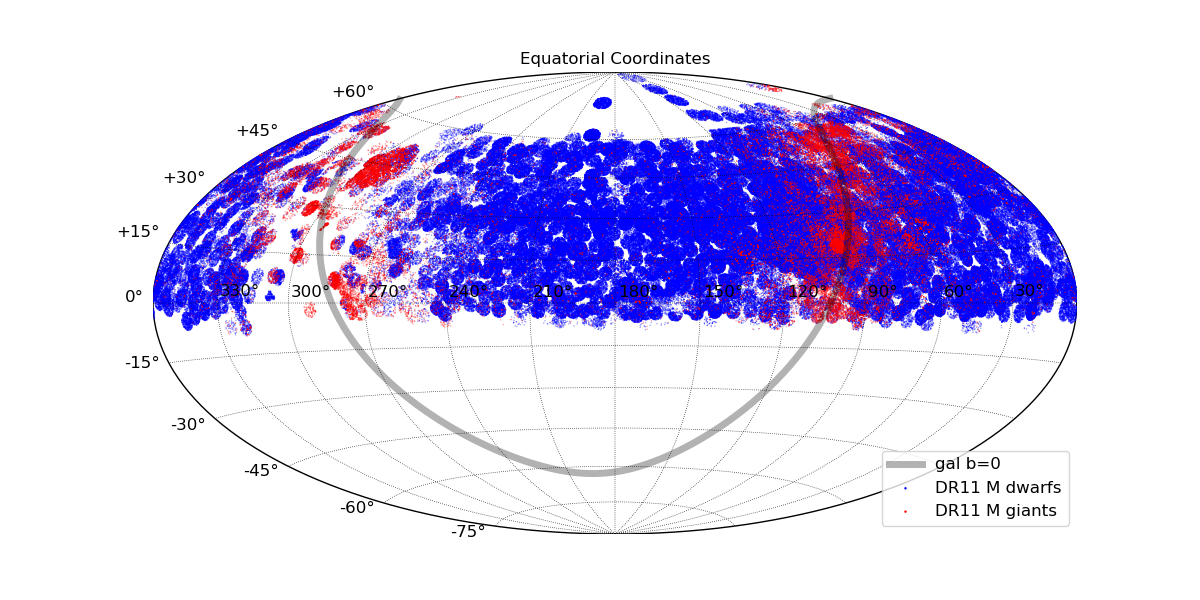}
	\caption{The sky distribution of the ‘footprints’ of the 671,654 M-type stars selected from LAMOST DR11, indicated by red for giants and blue for dwarfs. The map is projected using Equatorial Coordinates.}
  \label{fig:footprints}
\end{figure}

 Since the spectral features of cool stars are are most easily measured  in the $i$-band (6430-8630 \AA), we selected the LAMOST M-type samples in this work excluding the stars with a spectral  \textbf{S/N}  less than 10 in the $i$-band. Unless otherwise indicated, all the S/Ns presented hereafter refer to the S/N\_{$i$} value. From LAMOST DR11, we chose a total of 823,625 M-type spectra (S/N$\geq$10) representing 633,189 stars, which include 773,741 spectra corresponding to 594,564 M dwarfs and 49,884 spectra corresponding to 38,625 M giants. Figure \ref{fig:footprints} presents the spatial distribution of their 'footprints' in the sky. We observed that giants in the sample are generally found at low galactic latitudes and require larger reddening corrections compared to the dwarfs.

\subsection{Primary Samples}

In this work, our goal is to extract a pure subset from an extensive collection of observational data, concentrating on the types of stars that have many representatives in the solar neighborhood.  This selection bias might lead to the omission of uncommon stars, including metal-poor stars, subdwarfs, and young stellar objects,  among others. Our aim is to ensure that the parameters of typical stars are reliable and that uncommon types are not considered.

Our objective in choosing primary samples from the LAMOST catalog of 633,189 M-type stars is to pinpoint the typical stars in the color-magnitude diagram. We selected primary samples based on their locations in the absolute magnitude G versus the BP$-$RP diagram by cross-matching the LAMOST M-type catalog with Gaia EDR3 \citep{2021A&A...649A...1G}. The photometric data from Gaia are represented in the Vega magnitude system.  We excluded unreliable photometry and unresolved binary stars using the same criteria described in \cite{2024ApJS..270...32Q}. 

\begin{enumerate}
    \item Photometry error in G magnitude, BP magnitude, and RP magnitude is less than 0.08 mag;
    \item error\_Gmag/Gmag  $<$ 0.05, error\_BPmag/BPmag $<$ 0.05, and error\_RPmag/RPmag $<$ 0.05;
    \item Renormalised Unit Weight Error (RUWE) $<$ 1.4; and
    \item error\_parallax/parallax $<$ 0.2
\end{enumerate}

We excluded variable stars by cross-matching our sample with a list of variable stars collected from various surveys, including
Kepler \citep{2013MNRAS.432.1203M,2014ApJS..211...24M,2016AJ....151...68K,2019ApJS..244...21S}, ZTF \citep{2020ApJS..249...18C}, K2 \citep{2020yCat..36350043R}, WISE \citep{2018ApJS..237...28C}, Gaia \citep{2023A&A...674A...1G}, TESS \citep{2022RNAAS...6...96H,2022yCat..22580016P}, and LAMOST \citep{2022ApJS..259...11X}. We adopted the photogeometric
distances of \citet{2021AJ....161..147B}  as the distances of our sample, which was inferred from EDR3  parallaxes with a zero-point correction based on a three-dimensional Galactic model. The Pan-STARRS 1 3D dust map method \citep{2018MNRAS.478..651G}  was employed to correct the Gaia photometric data for extinction.

\subsubsection{M Dwarfs}
\begin{figure}[ht!]
	\centering
	\includegraphics[width=1.0\linewidth]{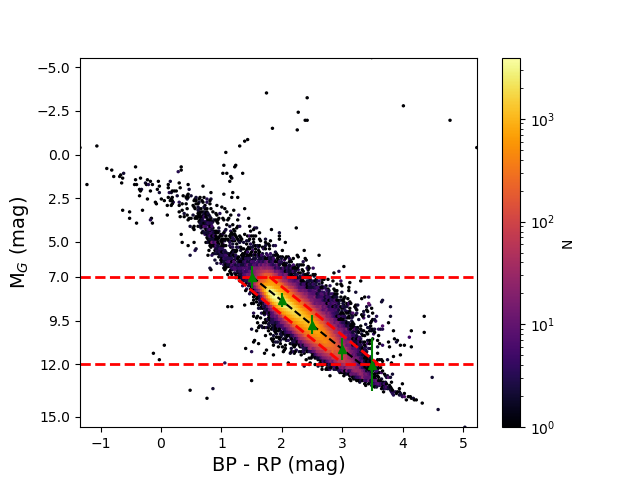}
	\caption{Color-magnitude diagram of the primary sample of M dwarfs. The error-bars represent the average values and the 1$\sigma$ uncertainties  of $G$ magnitudes with a color step of 0.5 mag, and the black dashed line is the first-order polynomial fit to the average values.  The oblique red dashed lines are created by shifting the black line up and down by  1$\sigma$ uncertainty on $G$ magnitude. The objects lie within the red dashed box are selected.       }
  \label{fig:HR-dwarfs}
\end{figure}

For M dwarfs, considering only those close to us can be observed  due to their low luminosities, we selected our sample with distances less than 2.0 kpc,  with typical distance uncertainties $<$ 2\% \citep{2021AJ....161..147B}. The $G$ magnitude limits for M dwarfs are set between 7--12 according to the contour shape in the color–magnitude diagram presented in Figure \ref{fig:HR-dwarfs}. To derive suitable dwarf samples, we calculated the average values and the 1$\sigma$ uncertainties of $G$ magnitudes using a color step of 0.5 mag. Subsequently, we performed a linear fit to the average values of $G$ magnitudes, indicated by the black dashed line in Figure \ref{fig:HR-dwarfs}. We shifted the black dashed line up and down by 1$\sigma$ uncertainty on $G$ magnitude (the average value of 1$\sigma$ uncertainties), which is 0.8 mag for dwarfs.  We obtained the dwarf sample based on the red dashed lines in the color–magnitude diagram shown in Figure \ref{fig:HR-dwarfs}. At this point, we selected 459,276 M dwarfs.

\subsubsection{M Giants}
\begin{figure}[ht!]
	\centering
	\includegraphics[width=1.0\linewidth]{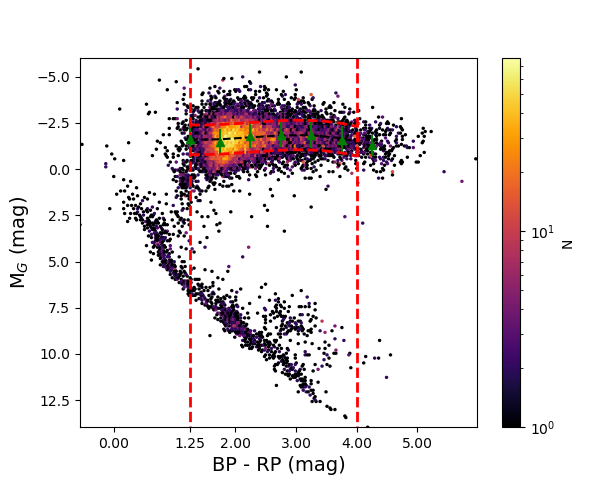}
	\caption{Color-magnitude diagram of the primary sample of M giants. The error-bars represent the average values and the 1$\sigma$ uncertainties  of $G$  magnitudes with a color step of 0.5 mag, and the black dashed curve is the third-order polynomial fit to the average values.  The red dashed curves are created by shifting the black curve up and down by 1$\sigma$ uncertainty on $G$ magnitude. The objects lie within the red dashed enclosed region are selected. }
  \label{fig:HR-giants}
\end{figure}

To maintain accuracy in the distance estimates for M giants, we excluded objects with distances greater than 5.0 kpc. For such circumstances, the chosen sample of giants typically shows a distance uncertainty of less than 4\% \citep{2021AJ....161..147B}. The color–magnitude diagram, illustrated in Figure \ref{fig:HR-giants}, shows the stellar locations of the M giants. Contaminants such as early-type stars, M dwarfs, and some white dwarfs are evident in the M-giant sample, as detailed in \citep{2023ApJS..266....4L}. We set color limits for M giants as BP-RP between 1.25--4.0 according to the contour shape in the color–magnitude diagram presented in Figure \ref{fig:HR-giants}. For giants, we also calculated the average values and the 1$\sigma$ uncertainties  of $G$  magnitudes with a color step of 0.5 mag. Then we performed a third-order polynomial fit to the the average values (the black dashed curve in Figure \ref{fig:HR-giants}). We shifted the black dashed curve up and down by 1$\sigma$ uncertainty on $G$ magnitude (0.8 mag).  We selected giant objects that lie within the red dashed enclosed region in the color-magnitude diagram presented in Figure \ref{fig:HR-giants}. After completing this process, a total of 13,796 M giants were chosen.

\subsubsection{Supplement}
\begin{figure}[ht!]
	\centering
	\includegraphics[width=1.0\linewidth]{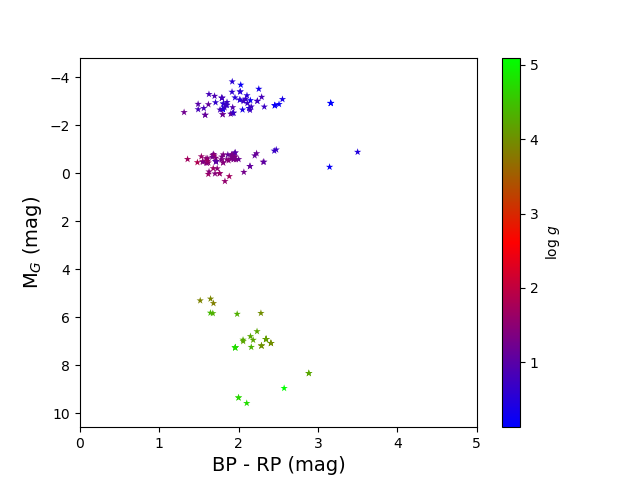}
	\caption{Color-magnitude diagram of the 160 added sample, color-coded by the log g values.}
  \label{fig:supplement}
\end{figure}

Furthermore, to complement the existing samples in parameter space, we incorporated 160 stars characterized by a LAMOST spectral S/N greater than 50, along with accurate parameter determinations from ASPCAP ($\sigma$\teff $<$ 10 K, $\sigma$\logg $<$ 0.05 dex, and $\sigma$[M/H] $<$ 0.02 dex).  Figure \ref{fig:supplement}
shows their locations on the Gaia color-magnitude diagram, which is color-coded by the log g values. The locations of giants and dwarfs align with their \logg\ values and are obvious in the color-magnitude diagram. Due to the high quality of their observed spectra and the precise parameter determinations, the 160 stars were incorporated into our primary samples. 

\subsection{Stellar Parameter Labels}

We labeled the spectra of the primary samples with stellar parameters collected from a list of catalogs,  and the  systematic discrepancies among these catalogs were corrected by aligning them with the StarHorse catalog \citep{2022A&A...658A..91A}.

\subsubsection{M Dwarfs}
\begin{figure*}[ht!]
	\centering
	\includegraphics[width=1.0\linewidth]{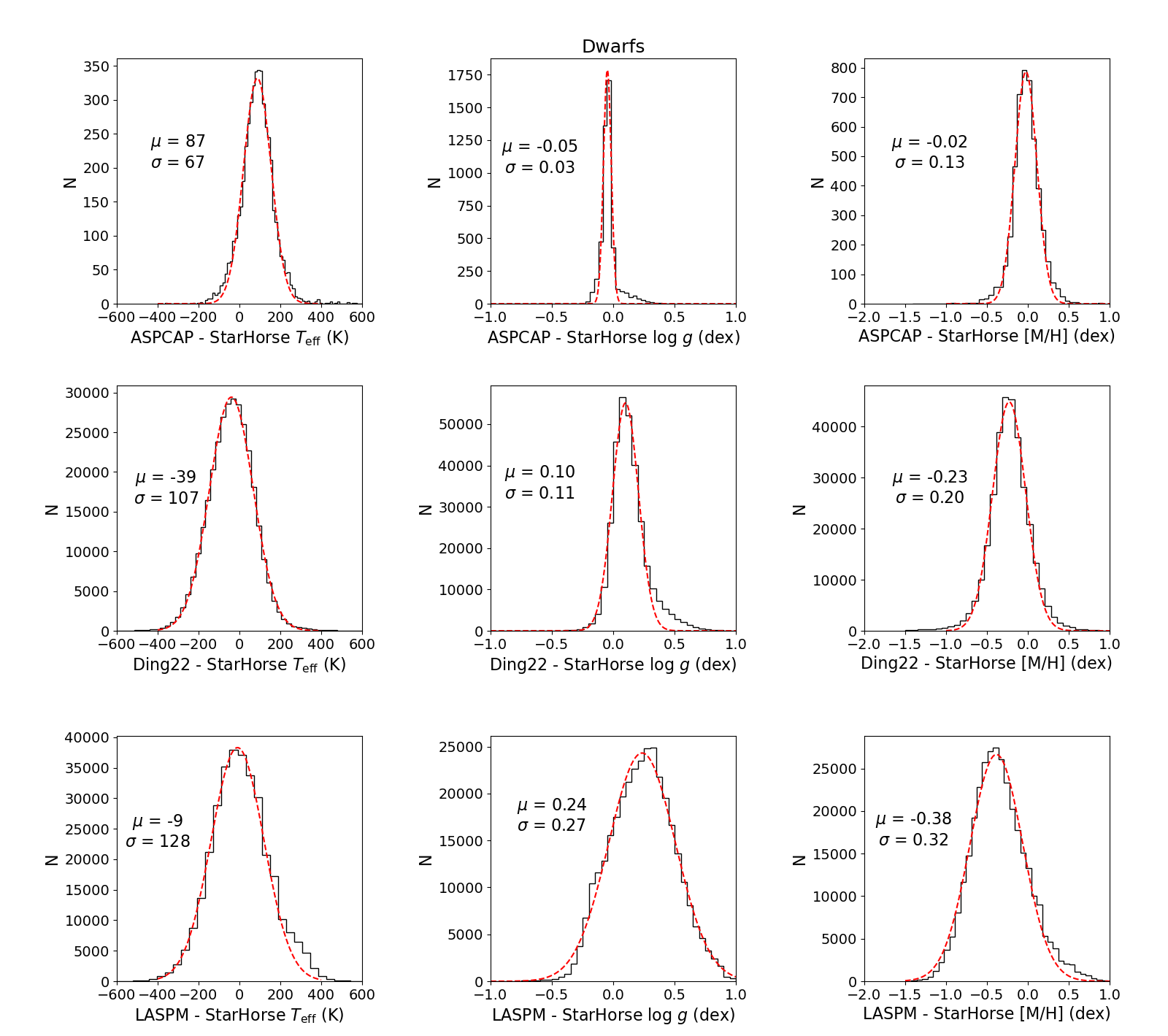}
	\caption{Histograms of differences between ASPCAP and StarHorse (top panel), Ding22 and StarHorse (middle panel), and LASPM and StarHorse (bottom panel). The red dashed curves are Gaussian fits to the distributions, and the mean and dispersion of the Gaussian fits to the mean and standard deviation values of the differences.}
 \label{fig:shift-dwarf}
\end{figure*}

For M dwarfs, we adopted stellar parameters from ASPCAP DR17, \cite{2022ApJS..260...45D} (hereafter referred to as Ding22),  and LASPM from pre-DR10. Figure \ref{fig:shift-dwarf} shows the differences in their respective comparisons with the StarHorse catalog. We found that there were varying offsets among the three catalogs compared to StarHorse.  The ASPCAP overestimated \teff \  by approximately 87 K, while there was no significant systematic difference for \logg \ and [M/H]. There was no significant shift between the \teff \ of LASPM and that of StarHorse, but large offsets for \logg \ and [M/H]. We shifted the stellar parameters of the three catalogs using their respective  systematic offsets relative to StarHorse.

 After applying the shifts derived from Figure \ref{fig:shift-dwarf}, we carried out the selection of parameter sample   based on the differences in parameters between each catalog and the StarHorse catalog.  We selected dwarf samples with parameter differences of  |$\Delta$\teff| $<$ 100 K, |$\Delta$\logg| $<$ 0.2 dex, and  |$\Delta$[M/H]| $<$ 0.2 dex,  given the standard deviation values of the differences presented in Figure \ref{fig:shift-dwarf}. In the process of parameter transfer, the ASPCAP catalog was prioritized above Ding22, and Ding22 was prioritized above LASPM. For example, when a star was parameterized by ASPCAP, Ding22, and LASPM, the parameters of ASPCAP were adopted, when parameterized by both Ding22 and LASPM, the parameters of Ding22 were adopted. At this point, we obtained stellar parameter labels for 41,990 M dwarf stars, with 3,384 stellar labels  from ASPCAP, 24,397 from Ding22, and 14,209 from LASPM.

\subsubsection{M Giants}
\begin{figure*}[ht!]
	\centering
	\includegraphics[width=1.0\linewidth]{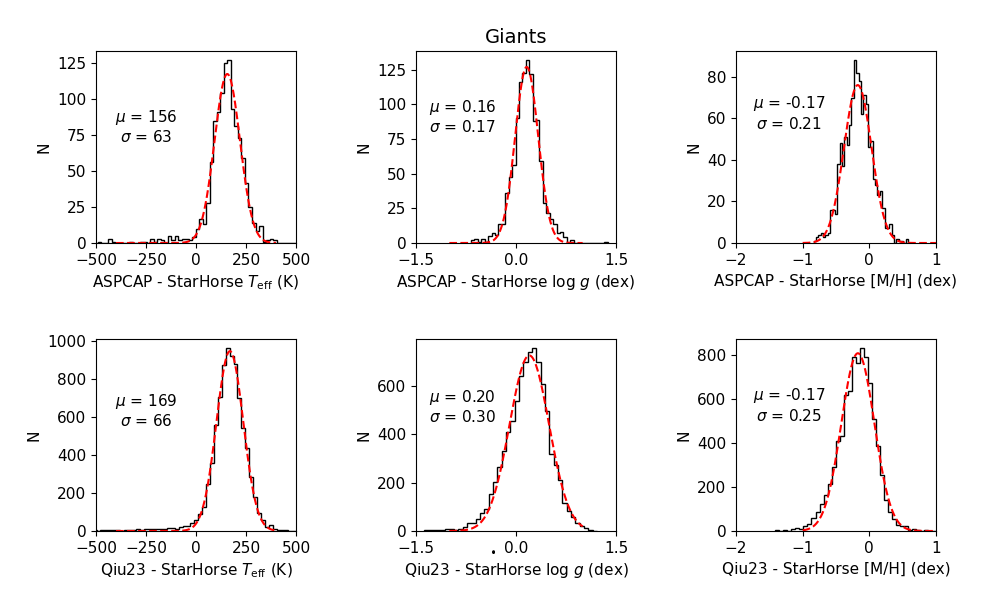}
	\caption{Histograms of differences between ASPCAP and StarHorse (top panel), and Qiu23 and StarHorse (bottom panel). The red dashed curves are Gaussian fits to the distributions, and the mean and dispersion of the Gaussian fits to the mean and standard deviation values of the differences.}
  \label{fig:shift-giant}
\end{figure*}

For M giants, we adopted stellar parameters from ASPCAP DR17 and \cite{2023RAA....23e5008Q} (hereafter referred to as  Qiu23). Figure \ref{fig:shift-giant} presents the differences between these catalogs and the StarHorse catalog. Since the Qiu23 parameters were transferred from those of ASPCAP, both catalogs have almost the same offsets relative to StarHorse. We still shifted the parameters of both catalogs, using their respective systematic offsets. Additionally, the parameter differences between each catalog and the StarHorse catalog were used to choose M giant samples. M giants were selected according to the criteria that the differences in the parameters |$\Delta$\teff| $<$ 70 K, |$\Delta$\logg| $<$ 0.2 dex, and |$\Delta$[M/H]| $<$ 0.2 dex, considering the standard deviations shown in Figure \ref{fig:shift-giant}.  In cases where parameter transfer was necessary, the ASPCAP catalog was prioritized over Qiu23. This implies that when a star was cataloged by both ASPCAP and Qiu23, the ASPCAP parameters were adopted. At this point, we obtained stellar parameter labels for 3,432 M giant stars, with 807 stellar labels from ASPCAP and 2,625 from Qiu23.

\subsection{Spectral Pre-processing}
\subsubsection{Back to Rest Frames}
For M-type stars, the RV of the LAMOST 1D pipeline was used to shift all of the spectra into their rest frames.  The precision quoted for the RV of the LAMOST 1D pipeline was approximately 5.0 \kms \citep{2015RAA....15.1095L,2021RAA....21..202D}.  The LAMOST low-resolution spectra were sampled in constant-velocity pixels, with a pixel scale of 69 \kms \ \citep{2019ApJS..240...10D}. The precision of the RV of the LAMOST 1D pipeline ($\sim$ 5.0 \kms) was less than 10\% of the pixel scale, which means that its RV calculations are accurate at subpixel values. Therefore, using the RV of the LAMOST 1D pipeline,  we completed an accurate shift to the rest frames.

\subsubsection{Flux Calibration and Dereddening} 

To compare the SEDs of stars with similar parameters,  it is necessary to correct the dereddening errors associated with the LAMOST flux calibration , particularly for the spectra of giant stars located at low galactic latitude. As described in \cite{2019ApJS..240...10D}, the LAMOST flux calibration introduces uncertainties to the SEDs of the calibrated spectra. Following the work done in \cite{2019ApJS..240...10D}, we recalibrated each spectrum of M-type stars by comparing the observed SED with the synthetic SED, using a second-order multiplicative polynomial to minimize recalibration impacts on the crowded bands of cool stars. For dwarfs, we adopted the BT-Settl model as the reference SED given the blue-band excess of the MARCS model, while for giants, we opted for the MARCS model over the BT-Settl model considering that the latter has a poor fit with the observed spectra for M giants.

Note that we adopted parameters aligned on StarHorse parameters and calibrated fluxes using the BT-Settl and MARCS SEDs. This approach may introduce a bias to the slope of the continuum by enforcing alignment with a different data set. However, the spectral feature details are minimally influenced by the flux calibration, and the parameter estimations primarily rely on these spectral feature details.

\subsection{Sample Cleaning}

It is not guaranteed that every sample obtained through automated selection would have reliable stellar labels and high-quality spectra. Furthermore, the data suffer from an imbalance issue, where most samples are clustered within certain parameter ranges.  Sample cleaning can remove samples with poor-quality spectra and, to some degree, also discard samples with unreliable stellar labels. In addition, the issue of data imbalance is addressed by regulating the number of samples within the parameter bins during sample cleaning. 

After the previous spectral pre-processing, we separated the spectra into the following parameter bins: \teff \ in steps of 50 K, \logg \ in steps of 0.1 dex, and [M/H] in steps of 0.1 dex. For groups with more than 150 spectra, only the first 150 spectra with the largest S/Ns were selected for clustering.  All of the spectra were resampled to a set of fixed wavelengths to align their wavelengths.

\begin{figure}[ht!]
	\centering
	\includegraphics[width=1.0\linewidth]{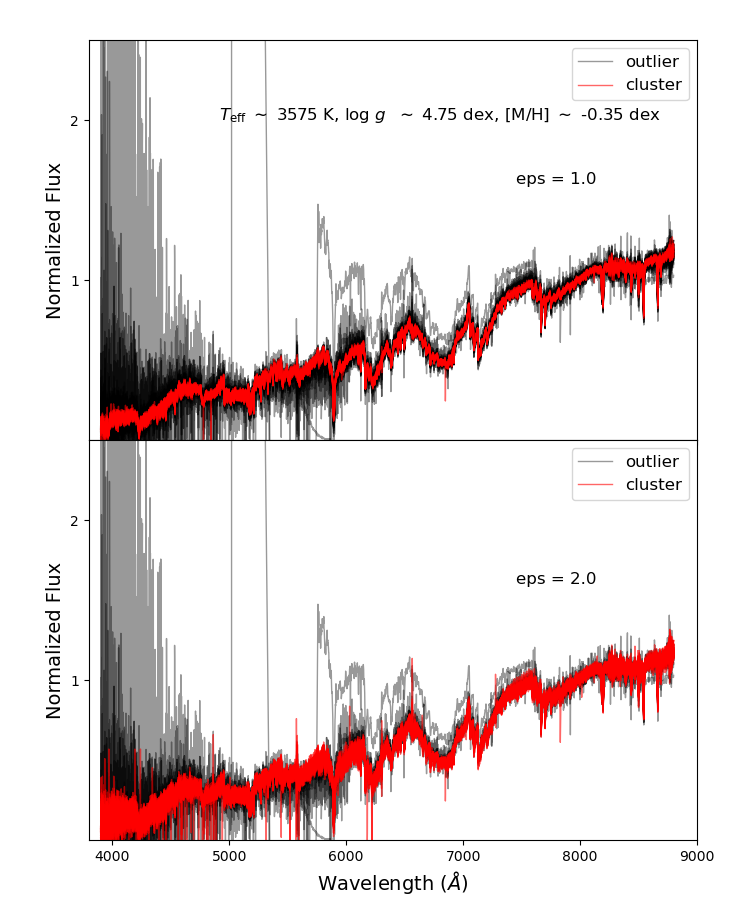}
	\caption{Examples of spectra clusterings for eps values of 1.0 (top panel) and 2.0 (bottom panel). The gray lines show all the spectra in the parameter bin of \teff: 3550--3600 K, \logg: 4.7--4.8 dex, and [M/H]: $-$0.3--$-$0.4 dex, while the red lines show the spectra that form a cluster.}
  \label{fig:cluster}
\end{figure}

We applied density-dased spatial clustering of applications with noise (DBSCAN) to the spectral clustering in each parameter bin. DBSCAN is robust to noise and can handle clusters of different densities and is also capable of identifying outliers. This makes it a good spectral cleaner for our samples.  DBSCAN requires two parameters: eps and minPts.  Here eps determines the radius around each point within which to search for neighboring points, and minPts is the minimum number of points required to form a cluster. Figure \ref{fig:cluster} displays examples of spectra clustering for eps values of 1.0  and 2.0. It is evident that as the eps value grows, the spectra differences  of the cluster increases, incorporating more spectra with lower S/N into the cluster.  We set eps$=$1.0 for spectral cleaning according to the calculated distances between the optical band spectra with a wavelength coverage of 3900--8800 \AA. We set different minPts values from 2 to 5 based on the number of samples in each parameter bin. 

In parameter bins with significantly more than five spectra, it is possible for a DBSCAN cluster to contain much more than  five spectra within the radius defined by eps. To prevent a disproportionate number of samples across various parameter bins, we selected a maximum of seven samples in a cluster  to ensure a balanced dataset suitable for machine learning.  For grids that contained only one sample or experienced clustering failure, we manually inspected their spectra for quality control.   We eventually obtained 5132 M-type stars after clustering cleaning and manual inspection.

\subsection{Further Cleaning Based on Neural Network}

To further cleanse and validate the self-consistency of stellar labels for the 5132 stars,  we trained a 5-layer neural network by randomly separating the 5132 stars into a training set and a testing set. The structure of \textbf{such} neural networks consists of a fully connected network with an input layer, three hidden layers, and an output layer.  Considering that LASPM only used the red part spectra, we also used the red part spectra (6000-8800 \AA) in neural network training. The spectral features were reduced from 1657 to 512, 512 to 64, then the 64 features were mapped to the output layer of 3 stellar labels.  

We conducted five separate training sessions for the neural networks, with each session consisting of 10,000 iterations. For each training session, we discarded outliers where the parameter differences between the predict results and the labeled parameters were |$\Delta$\teff| $\geq$ 100 K, |$\Delta$\logg| $\geq$ 0.2 dex, and |$\Delta$[M/H]| $\geq$ 0.2 dex. We concluded training  after the fifth training session because no outliers were detected in that round.

\begin{figure}[ht]
	\centering
	\includegraphics[width=1.0\linewidth]{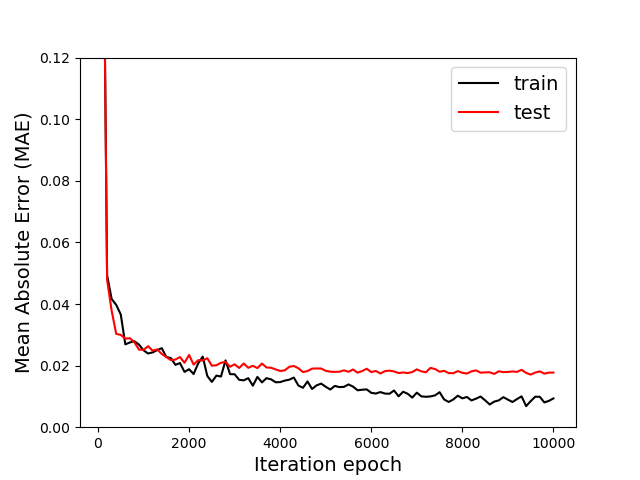}
	\caption{The loss function of the fifth training session with  a sampling interval of 100.}
  \label{fig:loss-function}
\end{figure}

\begin{figure*}[ht]
	\centering
	\includegraphics[width=0.8\linewidth]{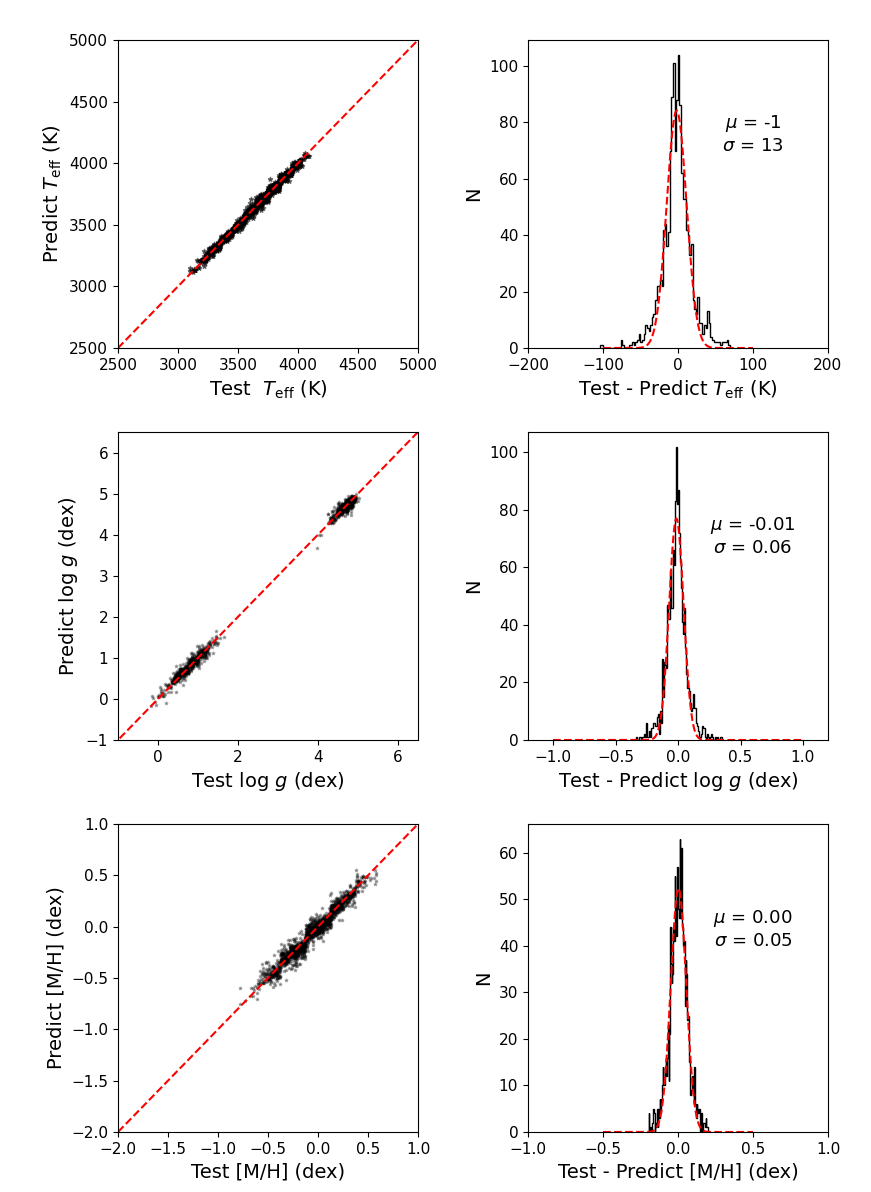}
	\caption{Comparison diagrams of the testing set between the true and predicted values for \teff (top panel), \logg(middle panel)  and [M/H] (bottom panel). The left panel shows one-to-one comparison diagrams and the right panel shows the histograms of differences.}
  \label{fig:test_comparison}
\end{figure*}

 We removed a total of 27 stars during this cleaning step. Figure \ref{fig:loss-function} presents the loss function of the  fifth training  session, indicating that there is no overfit.  Figure \ref{fig:test_comparison} shows the comparison diagrams of the testing set between the true and predicted values of the parameters. The predicted results were highly consistent with the labeled parameters, with a difference of 14 K for \teff, 0.06 dex for \logg, and 0.05 dex for [M/H]. This indicates that the parameter labels of the remaining 5105 M-type stars are reliable.

\subsection{Internal Cross-validation} \label{sec:cross-validation}
\begin{figure*}[ht!]
	\centering
	\includegraphics[width=0.8\linewidth]{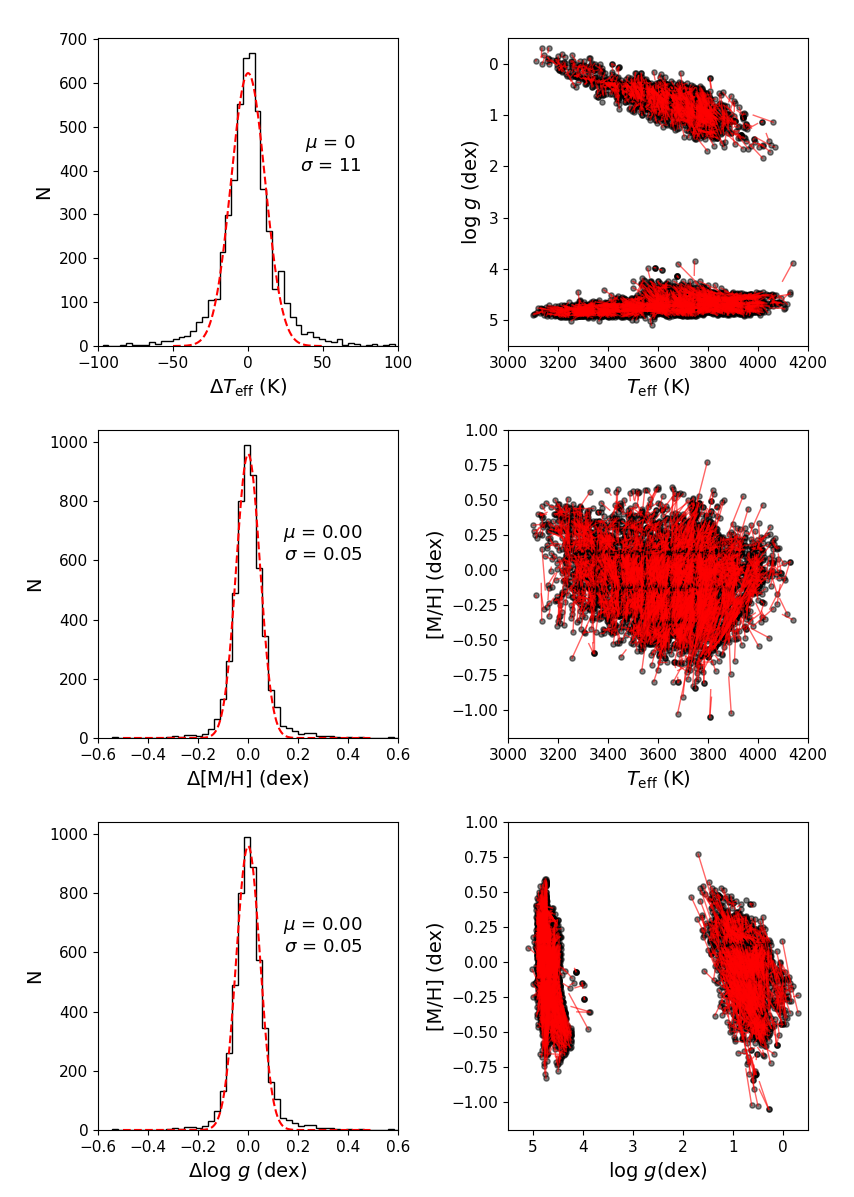}
	\caption{Left panel: histograms showing the differences between the derived parameters (\teff, [M/H], and \logg)  in the internal cross-validation process   and their stellar labels. Right panel: black points indicate the stellar labels, while red lines point to the derived parameters in the internal cross-validation process. }
  \label{fig:internal-cross-validation}
\end{figure*}

 To verify the reliability of the stellar labels of our samples, we performed an internal cross-validation.  We employed the template matching approach outlined in \citep{2021RAA....21..202D} to determine the parameters of  individual spectra from our samples. The approach interpolates between the parameters of the sample spectra by synthesizing linear combinations of the five best-agreement spectra. The template matching utilized the red part of the LAMOST spectra, ranging from 6000 to 8800 \AA.  We treated each spectrum in the samples as an unknown target and determined its parameters from the remaining samples. We compared these derived stellar parameters to their stellar labels. Figure \ref{fig:internal-cross-validation} illustrates the discrepancies between the derived parameters and their corresponding stellar labels, with a scatter of 11 K in \teff, 0.05 dex in [M/H], and 0.05 dex in \logg, indicating the accuracy of our sample's stellar labels.
 
Our samples span a finite parameter space, and it is essential for targets to match stars in the interior of its reference parameter space. Consequently, the derived parameters of stars at the edge of the parameter space were pulled toward the interior of the parameter distribution, as presented in the right panel of Figure \ref{fig:internal-cross-validation}. Fortunately, this interpolation impact is limited to the points located at the sparsely populated boundaries of the parameter space. For example, the metallicities of stars having a metallicity below -0.5 dex are overestimated due to interpolation, whereas stars with metallicities ranging from -0.5 to 0.5 dex remain unaffected by interpolation.
 
\clearpage

\section{The Results} \label{sec:results}
We ultimately obtained an empirical sample that included 5105 M-type spectra with homogeneous stellar labels through multiple cleanings.  Their stellar labels consisted of 881 labels from ASPCAP, 1395 from Qiu23, 1651 from Ding22, and 1178 from LASPM.  We packaged the stellar parameter catalog and the calibrated spectra in a FITS file, which is available online in the China-VO Paper Data Repository via\dataset[DOI: 10.12149/101488]{https://doi.org/10.12149/101488}. 

\subsection{The Stellar Parameter Coverage} \label{sec:coverage}
\begin{figure*}[ht!]
	\centering
	\includegraphics[width=1.0\linewidth]{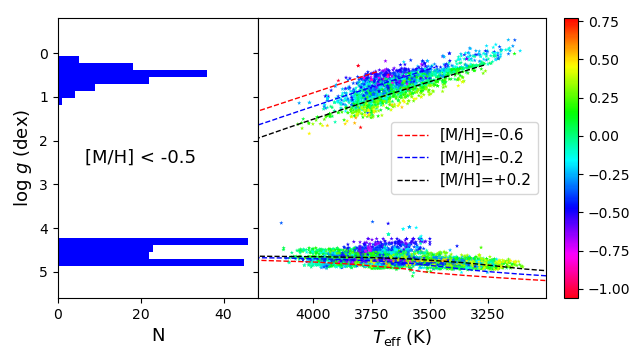}
	\caption{Right panel: the  Kiel diagram of the empirical sample. The color is coded by the  metallicity [M/H]. The dashed lines represent the isochrones from PARSEC model with different metallicities at the same age of 3 Gyr, i.e., $-$0.6, $-$0.2 and 0.2 dex for the red, blue and black lines, respectively. Left panel: histograms of \logg \ for [M/H] values less than -0.5 dex.}
  \label{fig:Kiel_sample}
\end{figure*}

Figure \ref{fig:Kiel_sample}. shows the Kiel diagram (\teff \ VS. \logg) for the 5105 M-type stars.  The PAdova and TRieste Stellar Evolution Code (PARSEC) isochrones at an age of 3 Gyr are also shown in Figure \ref{fig:Kiel_sample}. The different colors of the dashed lines represent different [M/H] values from PARSEC version 1.2S \citep{2012MNRAS.427..127B,2015MNRAS.452.1068C}. We notice that  our samples span a region of the Kiel diagram at \teff \ $\approx$ 3100--4150 K, \logg \ $\approx$ $-$0.3--5.1 dex, and [M/H] $\approx$  $-$1.0--0.7 dex. We can see separators on Kiel diagram to differentiate metallicities of giant stars, and the separators are consistent with the PARSEC isochrones. Separators to differentiate metallicities are not obvious for dwarf stars. The metal-poor dwarf stars ([M/H] $<$ -0.5 dex) display a double-sequence pattern (left panel of Figure \ref{fig:Kiel_sample}). A branch appears at approximately 4.3 dex in \logg, with all parameter labels derived from LASPM. It is hypothesized that this could be associated with the BT-Sett model, which is employed by both the LASPM and PARSEC models for M-type stars \citep{2014MNRAS.444.2525C}. We observed an identical pattern in the Kiel diagrams of StarHorse created by fitting PARSEC isochrones.   We notice a clear distinction of \logg \ between giants and dwarfs, and their locations on the Kiel diagram are consistent with the PARSEC theoretical tracks. 

To ensure the reliability of our samples, we limited our samples to  typical stars. This causes the parameter space coverage of our samples to be limited in the abundance patterns available within the solar neighborhood. We notice a scarcity of metal-poor stars, which means that the metallicity of a metal-poor star would be overestimated from our samples.

\subsection{The Spectra}
\begin{figure*}[ht!]
	\centering
	\includegraphics[width=1.0\linewidth]{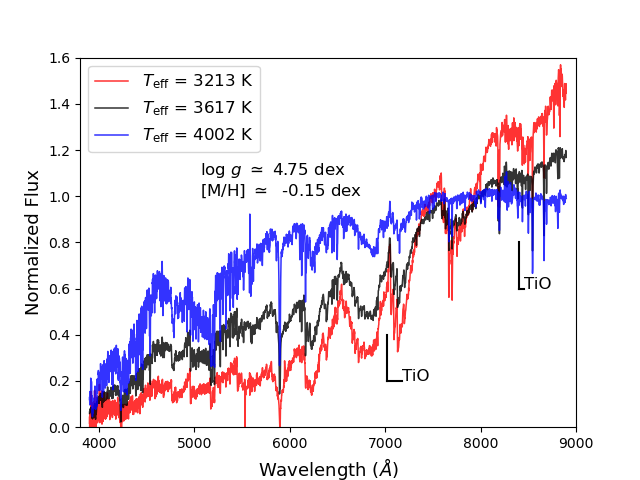}
	\caption{Effective temperature variations for the same surface gravity and Metallicity. The different colors of lines represent different \teff \ values. The \teff-sensitive TiO bands around 7050 and 8430 \AA \ are labeled.}
  \label{fig:teff_variations}
\end{figure*}

\textbf{The} collection of 5105 spectra is provided, housed within the first Header Data Unit (HDU) of the online FITS file. The second FITS table extension \textbf{contains} the parameters associated with the 5105 spectra.  Each spectrum features well-calibrated fluxes and wavelengths in the rest frame.  Figure \ref{fig:teff_variations} displays the spectral comparisons across various temperatures. The SED of the spectrum is notably affected by temperatures, with higher temperatures causing the more distinct features to appear in the blue region of the spectrum. In the red region of the spectrum, the TiO molecular bands near 7050 and 8430 \AA  \ exhibit a high sensitivity to \teff.

\begin{figure*}[ht!]
	\centering
	\includegraphics[width=0.8\linewidth]{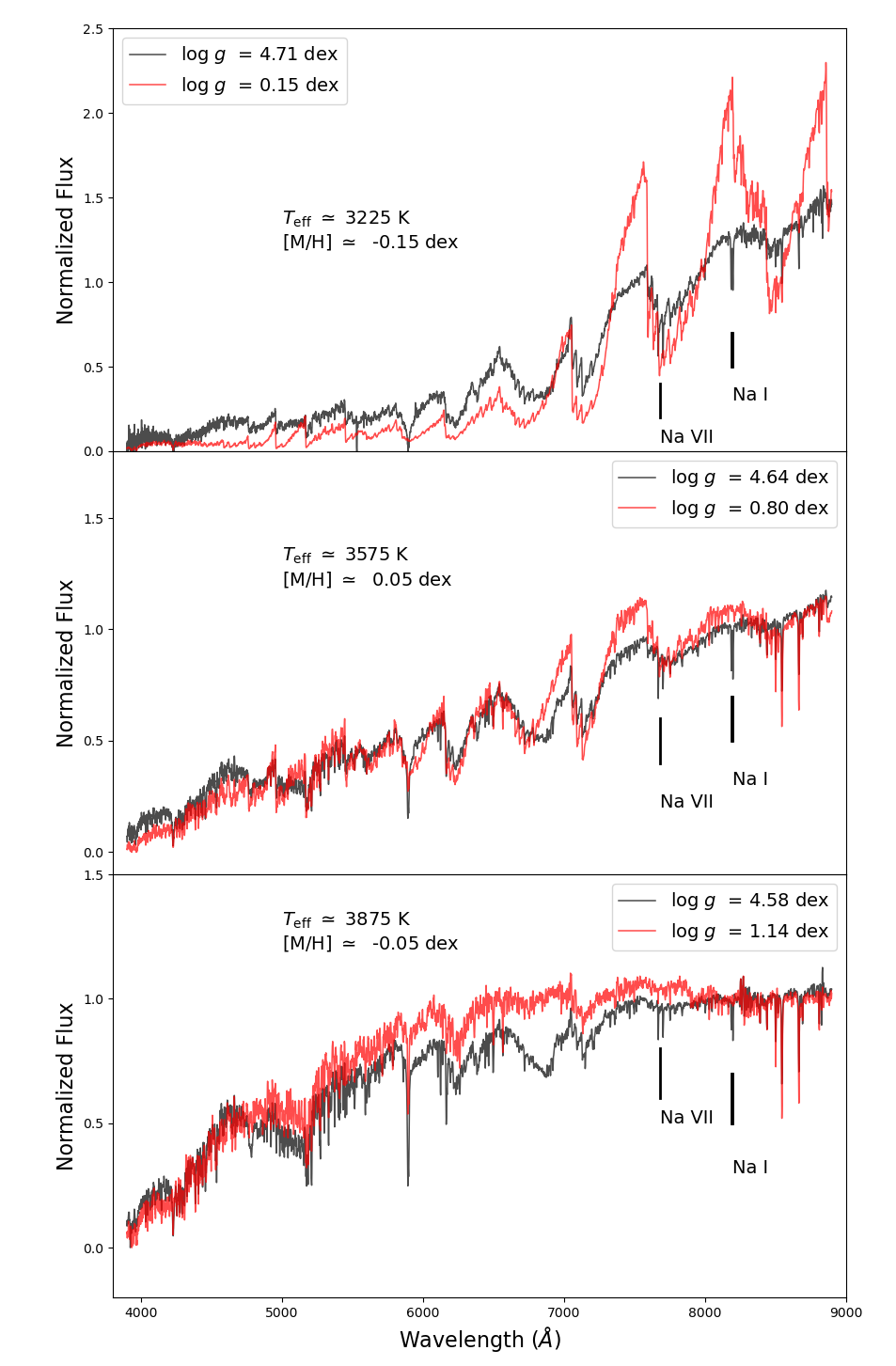}
	\caption{Surface gravity comparison between dwarf and giant spectra of the same metallicity and temperature for the entire spectrum. The black (red) line shows the dwarf (giant) spectrum. The Na- line pairs around 7680 \AA and 8190 \AA,  which are sensitive to \logg,   are labeled}
  \label{fig:logg_variations}
\end{figure*}

Figure \ref{fig:logg_variations} shows the spectral comparisons between dwarf and giant stars for a range of temperatures. Each pair of dwarf and giant stars exhibits identical metallicity and temperature. The molecular bands of giant and dwarf stars show a distinct shape difference in the red region of the spectrum ($\lambda$ $>$ 6700 \AA). A notable difference is observed between dwarf and giant stars around 6700 \AA, which is due to CaH, an effective indicator of luminosity \citep{2009ssc..book.....G}. The Na- line pairs  around 7680 \AA \ and 8190 \AA \ exhibit significant changes in their line wings due to pressure broadening, making them good gravity indicators. 

\begin{figure*}[ht!]
	\centering
	\includegraphics[width=0.8\linewidth]{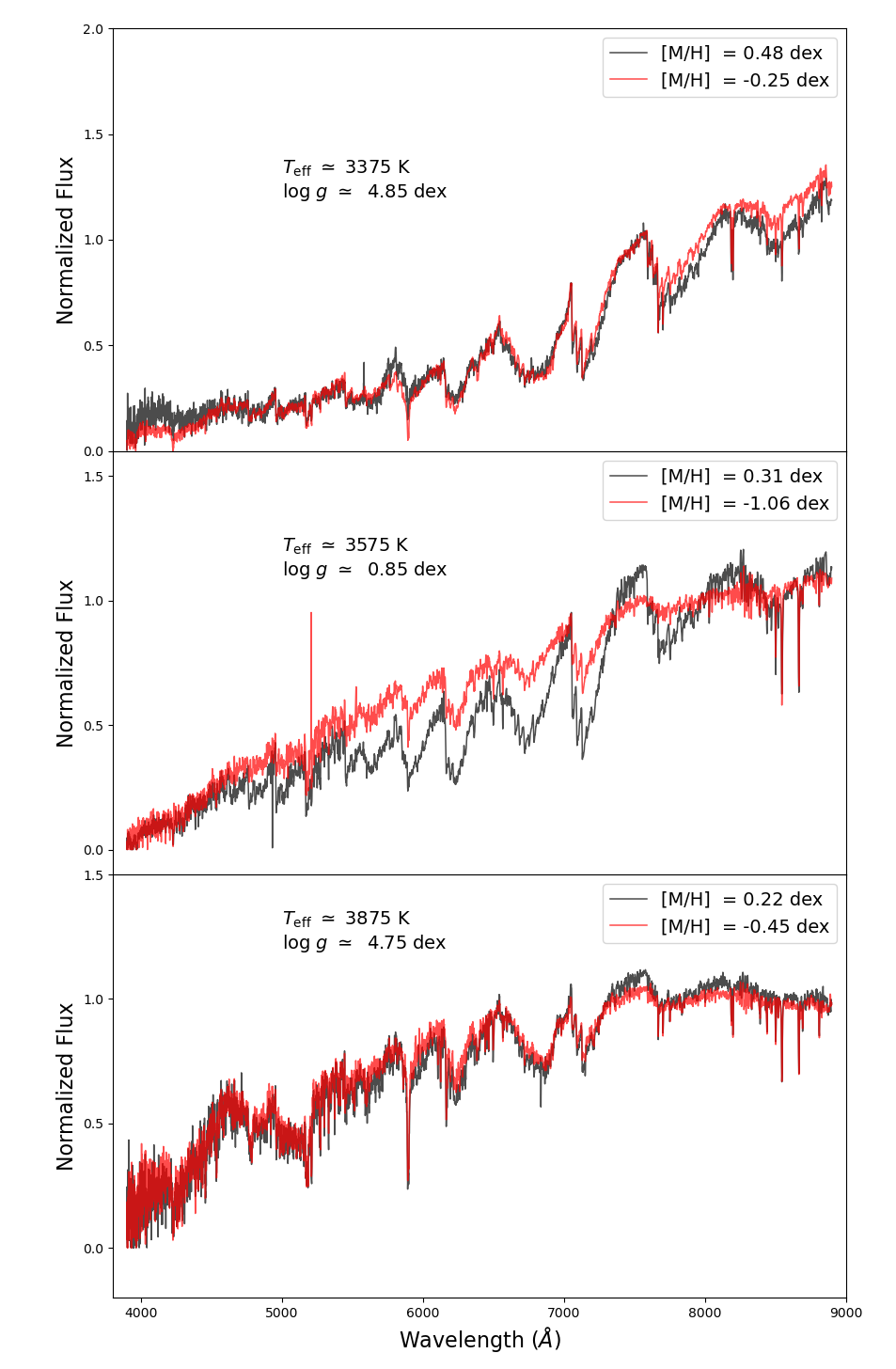}
	\caption{Metallicity comparison between spectra of the same surface gravity and temperature for the entire spectrum.The different colors of lines represent different [M/H] values. }
  \label{fig:metallicity_variations}
\end{figure*}

Figure \ref{fig:metallicity_variations} shows the differences in the spectra between different metallicities. Each pair of spectra exhibits identical surface gravity and temperature. It is observed that the TiO bands vary with metallicity. This alteration in TiO bands is more susceptible to flux calibration compared to stars of earlier types, which results in the M-type metallicity being more challenging to determine compared to \teff \ and \logg.

\clearpage

\section{Application to LASPM}  \label{sec:application}

The work of the previous sections was used to update LASPM and to determine the stellar parameters for 877,570 M-type spectra (corresponding to 667,483 stars) in LAMOST DR11. Rather than utilizing all 5105 spectra, we separated the 5105 stars into 1365 parameter bins (50 K, 0.1 dex, 0.1 dex) and chose the spectrum with the highest S/N from each bin to serve as our reference set.  The reference data set is saved in an FITS file, where the spectra are located in the primary HDU and the associated parameters are in the second extension, accessible for online download via\dataset[DOI: 10.12149/101488]{https://doi.org/10.12149/101488}. We  utilized the template matching approach described in \citep{2021RAA....21..202D},  which interpolates between the reference parameters by synthesizing linear combinations of the five best-agreement spectra. The template matching used the red part of the LAMOST spectra, ranging from 6000 to 8800 \AA.

\subsection{The Kiel Diagrams}

We provide the stellar atmospheric parameter catalog of the updated LASPM, which includes 877,570 spectra corresponding to 667,483 M-type stars from LAMOST \textbf{DR10 and} DR11. In \textbf{Table} \ref{tab:parameters}, we display a
sample set of our stellar parameter results. The attribute column is not fully presented here. Additional significant attributes, including spectral quality details such as S/N in the g and r bands, as well as spectral identification information such as the LAMOST designation and the unique spectra ID (specid), are available in the full catalog, accessible online via\dataset[DOI: 10.12149/101488]{https://doi.org/10.12149/101488}. Additionally, this catalog of stellar parameters can be found on the LAMOST's official data release  website\footnote{\url{https://www.lamost.org/dr10/v2.0/}\label{fn:dr10-release}}.

\begin{deluxetable*}{ccc ccc ccc ccc c}[ht]
	\tablewidth{700pt}
	\tabletypesize{\scriptsize}
	\tablecaption{Stellar Parameters of Randomly Selected stars. 
		\label{tab:parameters} }
	\tablehead{
		\colhead{source\_id$^a$} & \colhead{R.A.} & \colhead{Dec.} &\colhead{\teff} & \colhead{error\_\teff} & \colhead{\logg} & \colhead{error\_\logg} & \colhead{[M/H]} &  \colhead{error\_[M/H]} & \colhead{RV} & \colhead{error\_RV} & \colhead{S/N\_i} \\
		\colhead{ } & \multicolumn{2}{c}{(deg)} & \multicolumn{2}{c}{(K)} & \multicolumn{2}{c}{(dex)} & \multicolumn{2}{c}{(dex)} &\multicolumn{2}{c}{(\kms)} & \colhead{ } }
	
    \startdata 
	3023924638746047744&86.8664970&-3.7075162&3491&78&4.74&0.09&-0.01&0.16 &34.5&7.3&74.50 \\
    3309400536408261248&67.9177800&14.5833930&3588&55&4.77&0.06&0.07&0.10&-36.7&4.1&26.15 \\
    3947992246959480064&189.6793780&18.5449010&3486,&71&0.39&0.13&-0.37&0.16&100.2&5.9&275.01\\
    1325838816385675008&247.0077780&33.5193720&3834&23&4.62&0.03&-0.04&0.05&-44.8&3.0&55.90\\
    \\ \hline
    \enddata    
\tablecomments{The attribute column is not fully presented here.The full catalog can be accessed in a machine-readable format in the online journal.\\
            $^a$: The source identiﬁer (source\_id) from Gaia EDR3\\
               }
\end{deluxetable*}

\begin{figure}[ht!]
	\centering
	\includegraphics[width=1.0\linewidth]{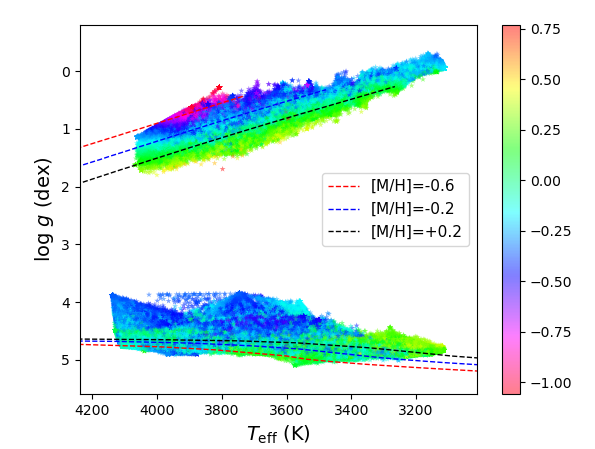}
	\caption{ The LASPM-derived Kiel diagram of LAMOST DR11. The color is coded by the  metallicity [M/H]. The dashed lines represent the isochrones from PARSEC model with different metallicities at the same age of 3 Gyr, i.e., $-$0.6, $-$0.2 and 0.2 dex for the red, blue and black lines, respectively.}
  \label{fig:Kiel_laspm}
\end{figure}

Figure \ref{fig:Kiel_laspm} shows the Kiel diagram for the M-type stars of LAMOST DR11. We also show the PARSEC isochrones as in Figure \ref{fig:Kiel_sample}. The stellar parameter distribution of all the M-type stars of DR11 is basically similar to that of the 5105 samples, as expected.  

\begin{figure}[ht!]
	\centering
	\includegraphics[width=1.0\linewidth]{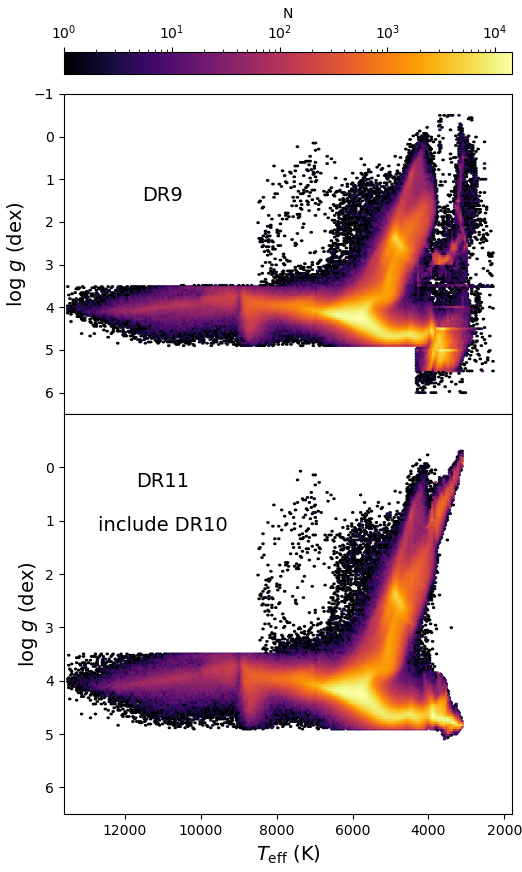}
	\caption{Kiel diagram of all stars of LAMOST DR9 (upper) and DR11 (lower), color coded by the density. The stellar parameter catalog for M-type stars has been updated in DR10, which is a subset of DR11.}
  \label{fig:Kiel_lasp}
\end{figure}

Figure \ref{fig:Kiel_lasp} shows the Kiel diagram for all stars of LAMOST DR11. The stellar parameter catalog for M-type stars has been updated in DR10, which is a subset of DR11. For comparison, the Kiel diagram of DR9 is also presented.  For AFGK-type stars \textbf{in all data releases}, the stellar parameters were determined by LASP through comparing the LAMOST spectra to the ELODIE library, which is documented in \cite{2015RAA....15.1095L} and \cite{2019ApJS..240...10D}. The LASPA (LASP for A-type stars)  used different spectral features with those used by LASPFGK (LASP for FGK-type stars), and masked the Ca II HK (3900--4060 \AA), H$\beta$ (4857--4867 \AA), and H$\alpha$ (6400--6800 \AA) to reduce the effect of the features of Am and Ae stars on the parameter  estimates (\textbf{consult the online documentation via the link\footnote{\url{https://www.lamost.org/dr10/v2.0/doc/lr-data-production-description}}}). Therefore, we can see a small gap around \teff \ $\sim$ 8500 K on the Kiel diagram. Thanks to the StarHorse catalog as a benchmark to correct the systematic differences between different spectroscopic measurements, we ultimately obtained a basically continuous Kiel distribution diagram, marking a notable improvement compared to DR9.

Due to the limited metallicity coverage of our samples, M giants contain fewer metal-poor giants compared to K giants, resulting in M giants occupying a narrower region on the Kiel diagram compared to K giants. The LASP depends on the classification given by the LAMOST 1D-pipeline to select the suitable pipeline for parameter estimation. For instance, in cases where a spectrum is recognized to be M-type, the LASPM is used for analysis, whereas LASPFGK is employed when the spectrum is classified as K-type. The classification accuracy of the LAMOST 1D pipeline varies by approximately 5 subtypes \citep{2015RAA....15.1095L}, indicating that spectra which are late K-type could be misclassified as early M-type, while those that are early M-type might be mistaken for K-type spectra. LASPM and LASPFGK utilize distinct spectral bands and varied references, resulting in the kinks and sequence-split observed in the lower main sequence around \teff \ $\sim$ 4000 K.

\subsection{Precision of the updated LASPM}
We calculated the parameter precision from the parameter estimates of repeated observations for the same stars. We selected M-type stars with each star having at least three observations.  We used an unbiased estimator defined in \cite{2021RAA....21..202D},  to measure precision. The estimator is given below:
\begin{equation}
\epsilon =  \sqrt{\frac{\rm N}{\rm N-1}} \times ({\rm P_{\rm i}} - \overline{\rm P})
\label{eq:eq1}
\end{equation}

where N is the number of repeated observations, ${\rm P_{\rm i}}$ is the parameter in terms of \teff, \logg, and [M/H]  of the ${\rm i_{\rm th}}$ observation, and $\overline{\rm P}$ = $\frac{1}{\rm N}\sum_{\rm i}^{\rm N} {\rm P_{\rm i}}$. 

\begin{figure}[ht!]
	\centering
	\includegraphics[width=1.0\linewidth]{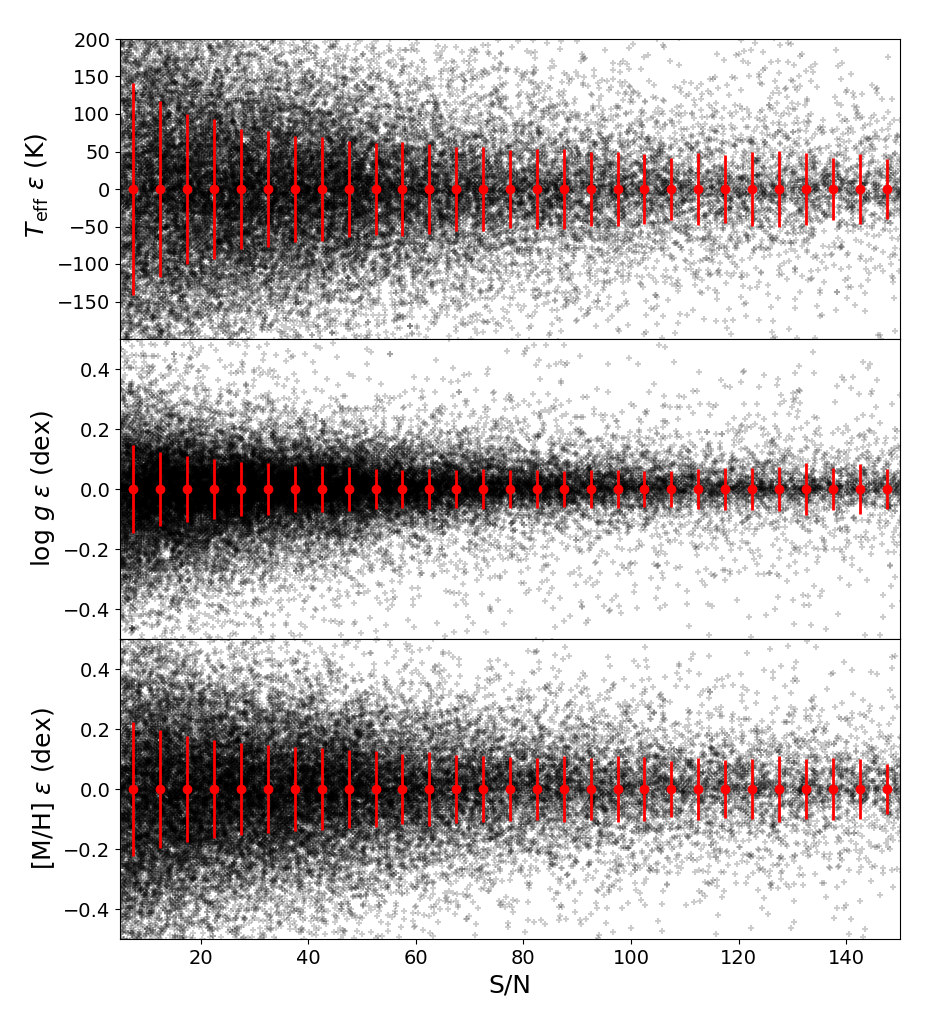}
	\caption{The parameter errors at different S/Ns in terms of \teff \ (top), \logg \  (middle), and [M/H] (bottow). The error bars in red  represent the parameter errors in different  S/N bins, with a step of 5.0.}
  \label{fig:precision}
\end{figure}

\begin{figure*}[ht!]
	\centering
	\includegraphics[width=1.0\linewidth]{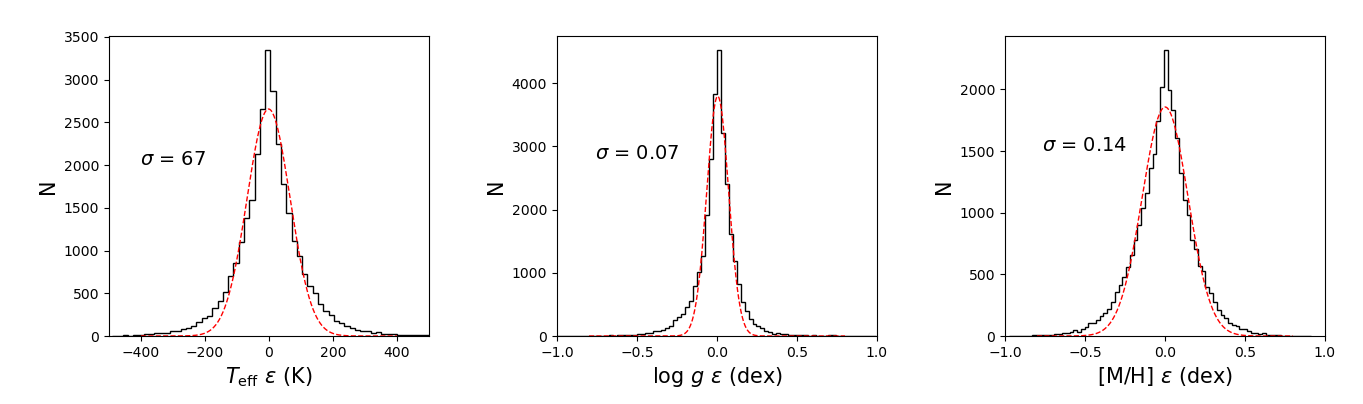}
	\caption{Histograms of $\epsilon$ for \teff \ (left), \logg \ (middle), and [M/H] (right). The unbiased estimator, $\epsilon$, was derived from parameter estimates of multi observations with spectral S/N greater than 10.0 for the same M-type stars.  The histogram is fitted by Gaussian shown in red dashed curves; The parameter precision (1$\sigma$ uncertainty) for the three parameters are labeled. }
  \label{fig:hist_precision}
\end{figure*}

Figure \ref{fig:precision} shows the variation of the parameter errors with the spectral S/N. We notice that the parameter errors have a clear increase with decreasing of S/N when S/N $<$ 20, while when S/N $>$ 40, the error is almost constant at |$\Delta$\teff| $\approx$ 60 K, |$\Delta$\logg | $\approx$ 0.10 dex  and |$\Delta$[M/H]| $\approx$ 0.15 dex.  Figure \ref{fig:hist_precision} shows the Gaussian fits to the $\epsilon$ histograms of \teff, \logg \ and [M/H]  derived from repeated observations with spectral S/N $\geq$ 10. The 1$\sigma$ uncertainties of the $\epsilon$ distributions are 67 K for \teff, 0.07 dex for \logg, and 0.14 dex for [M/H], respectively.  The precision (in terms of 1$\sigma$ uncertainties) of the updated LASPM is improved over its predecessor,  which achieved accuracies of 118 K for \teff, 0.20 dex for \logg,  and 0.29 dex for [M/H]. This arises because the reference dataset comprises a dense grid of data that closely mirrors the natural distribution of observations.

\subsection{Comparison with StarHorse}
\begin{figure*}[ht!]
	\centering
	\includegraphics[width=0.8\linewidth]{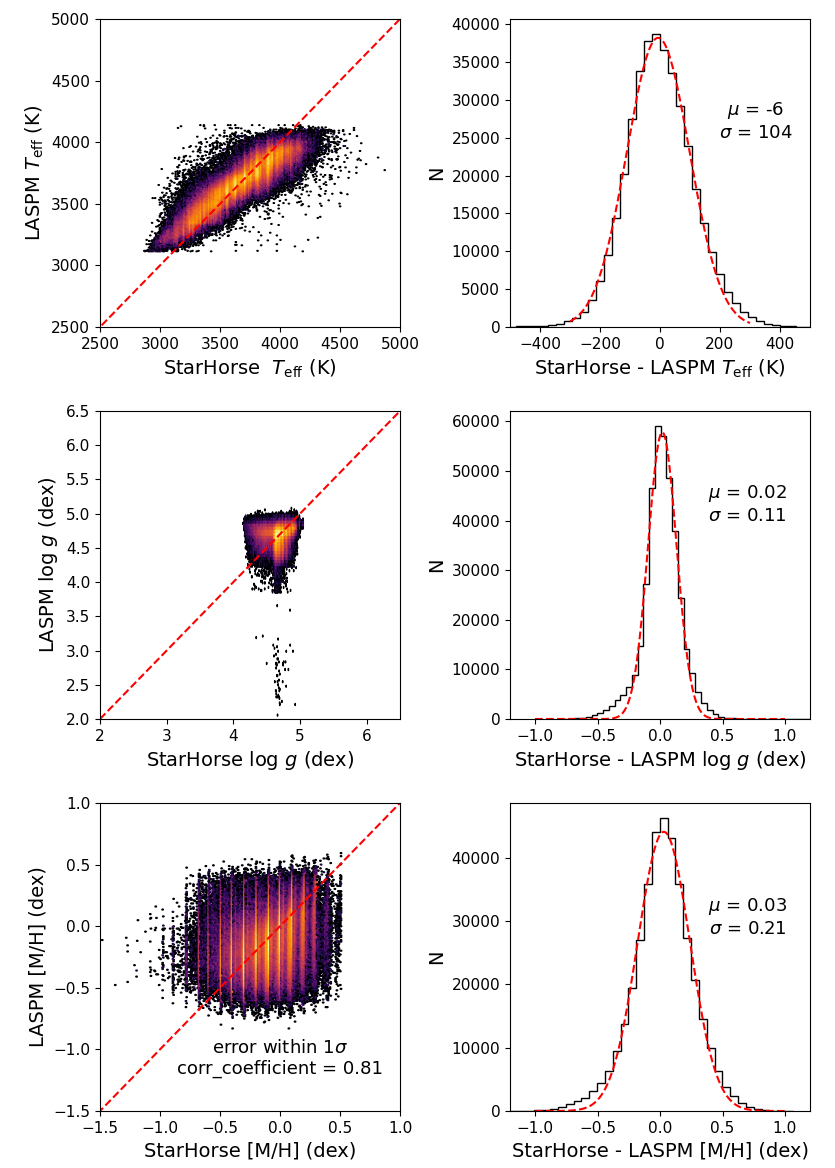}
	\caption{The left panel shows density plots comparing the LASPM results  with those from StarHorse for M dwarf stars.The right panel shows histograms of differences: no systematic shift was noticed. The correlation coefficient between LASPM and StarHorse [M/H], omitting data points with errors exceeding one standard deviation (1$\sigma$), is presented. }
  \label{fig:starhorse_dwarfs}
\end{figure*}

\begin{figure*}[ht!]
	\centering
	\includegraphics[width=0.8\linewidth]{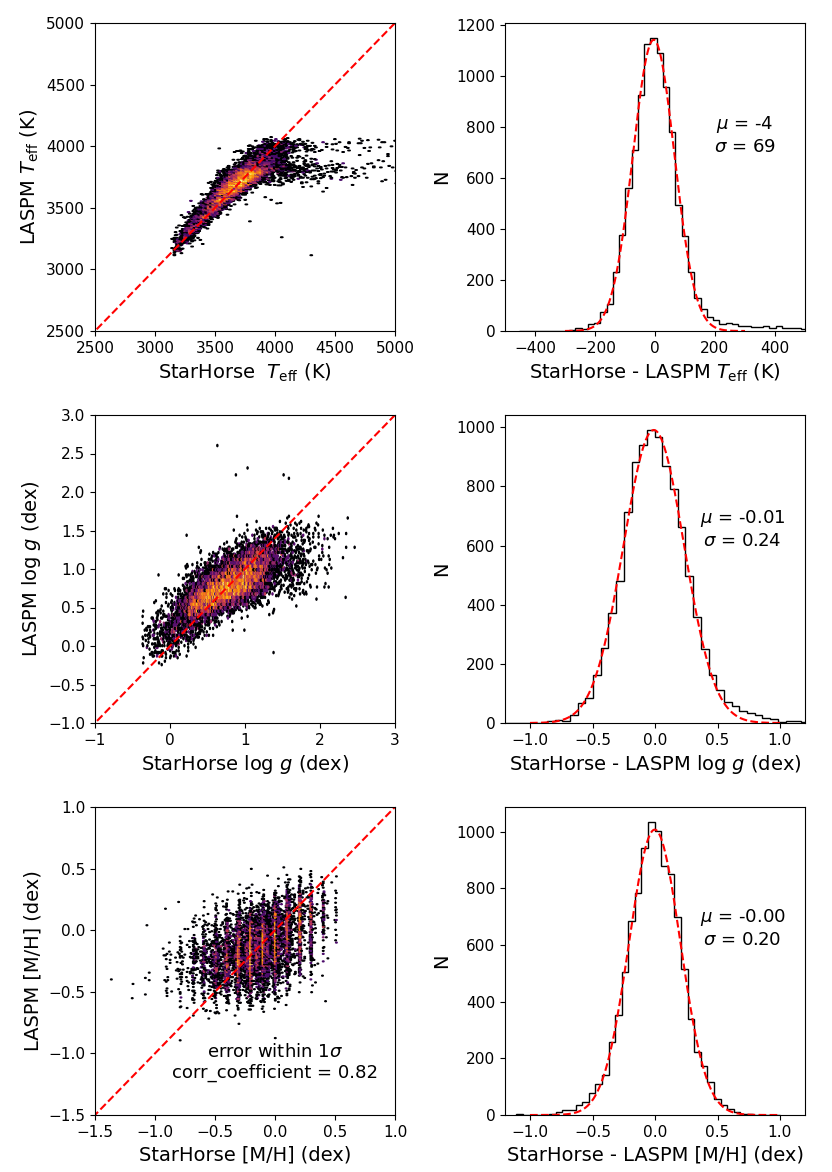}
	\caption{The left panel shows density plots comparing the LASPM results with those from StarHorse for M giant stars.The right panel shows histograms of differences: no systematic shift was noticed. The correlation coefficient between LASPM and StarHorse [M/H], omitting data points with errors exceeding one standard deviation (1$\sigma$), is presented.}
  \label{fig:starhorse_giants}
\end{figure*}

To investigate whether the revised LASPM algorithm induces systematic errors in parameter estimates, we compared the LASPM results to the StarHorse catalog. The LASPM parameter catalog was cross-matched with the StarHorse catalog, resulting in a match of 392,509 stars with the LAMOST spectral S/N $>$ 10. This dataset includes 382,550 M dwarf stars  and 9,959 M giant stars. Figure \ref{fig:starhorse_dwarfs} shows the comparison  \textbf{between}  stellar parameters derived from LASPM  \textbf{and} StarHorse for M dwarf stars, whereas Figure \ref{fig:starhorse_giants} presents the comparison for M giant stars. The parameters of LASPM are consistent with those of StarHorse for both M dwarf and M giant stars, showing a scatter of  104 (69) K in 
 \teff,  0.11 (0.24) dex in \logg, and 0.21 (0.20) dex in [M/H],  for M dwarf (giant) stars. The biases of 6 (4) K in \teff, 0.02 (0.01) dex in \logg, and 0.03 (0.00) dex in [M/H], for M dwarf (giant) stars,  are  significantly smaller than the respective scatters, suggesting the absence of systematic offset. 
 
 For M dwarf stars, we observed a general pattern in \teff\ where the higher temperatures tend to be underestimated and the lower temperatures tend to be overestimated. This phenomenon can be attributed to the limited temperature range covered by our reference samples, which spans from 3100 K to 4150 K, and target stars are required to match stars within that temperature range. This also results in an underestimation of the temperatures for giant stars exceeding 4000K. Similarly, the trend for metallicity shows that [M/H] values below -0.5 dex tend to be overestimated due to interpolation within the metallicity range of our \textbf{reference} samples.  The stripe-like overdensities in the bottom left plot correspond to the metallicity resolution of the stellar-model grid that StarHorse used. We calculated the correlation coefficient between LASPM and StarHorse [M/H], excluding data points with errors exceeding one standard deviation (1$\sigma$).  The metallicity correlation coefficients for both dwarf and giant stars are greater than 0.80, indicating that most of data points are concentrated around the one-to-one line.  Obtaining accurate metallicity only from photometric data is unlikely to be achieved, which accounts for the considerable variation in metallicity values seen in Figure \ref{fig:starhorse_dwarfs} and Figure \ref{fig:starhorse_giants}.


\subsection{Comparison with APOGEE}
We cross-matched the LASPM catalog with the APOGEE DR17 following these criteria:
\begin{enumerate}
\item The APOGEE STARFLAG $=$ 0;
\item The LAMOST DR11 spectra are identified as M type through the LAMOST 1D pipeline and are parameterized by LASPM;
\item The S/N of the $i$ band for the LAMOST spectra should be larger than 10.
\end{enumerate}

In this way, we selected 12 178 spectra of 6552 M dwarf stars and 3304 spectra of 2068 M giant stars after cross-matching.  We adjusted the APOGEE stellar parameters by applying the same offsets between ASPCAP and StarHorse to both dwarfs and giants, respectively. Refer to Figure \ref{fig:shift-dwarf} and Figure \ref{fig:shift-giant}  for detailed offset values.

\begin{figure*}[ht!]
	\centering
	\includegraphics[width=0.8\linewidth]{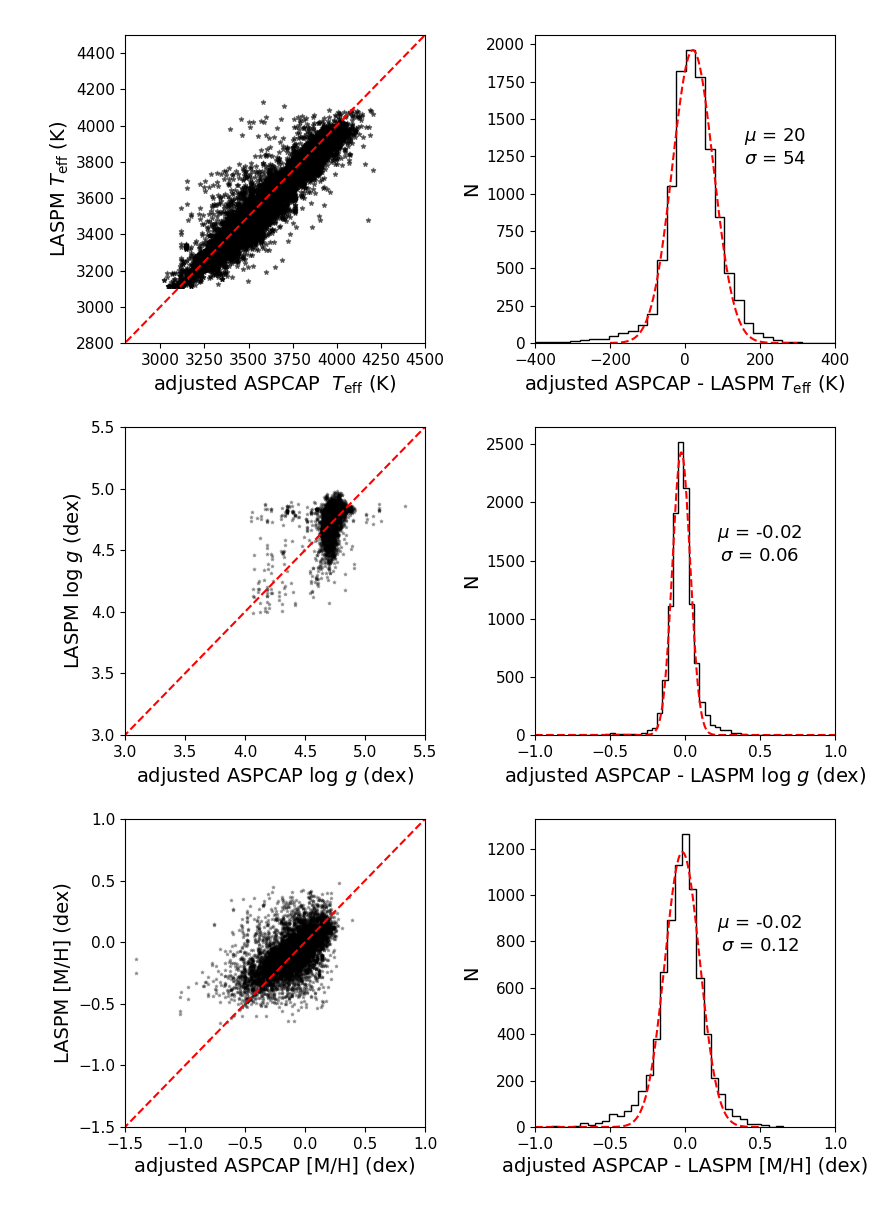}
	\caption{Comparison of the stellar parameters derived from LASPM to the adjusted parameters of ASPCAP for M dwarfs,  where adjusted\_\teff \ = teff - 87 (K), adjusted\_\logg \ = \logg \ + 0.05 (dex), and adjusted\_[M/H] = [M/H] + 0.02 (dex). The left panel shows one-to-one comparison diagrams and the right panel shows the histograms of differences.}
  \label{fig:compare_APOGEE_dwarf}
\end{figure*}

\begin{figure*}[ht!]
	\centering
	\includegraphics[width=0.8\linewidth]{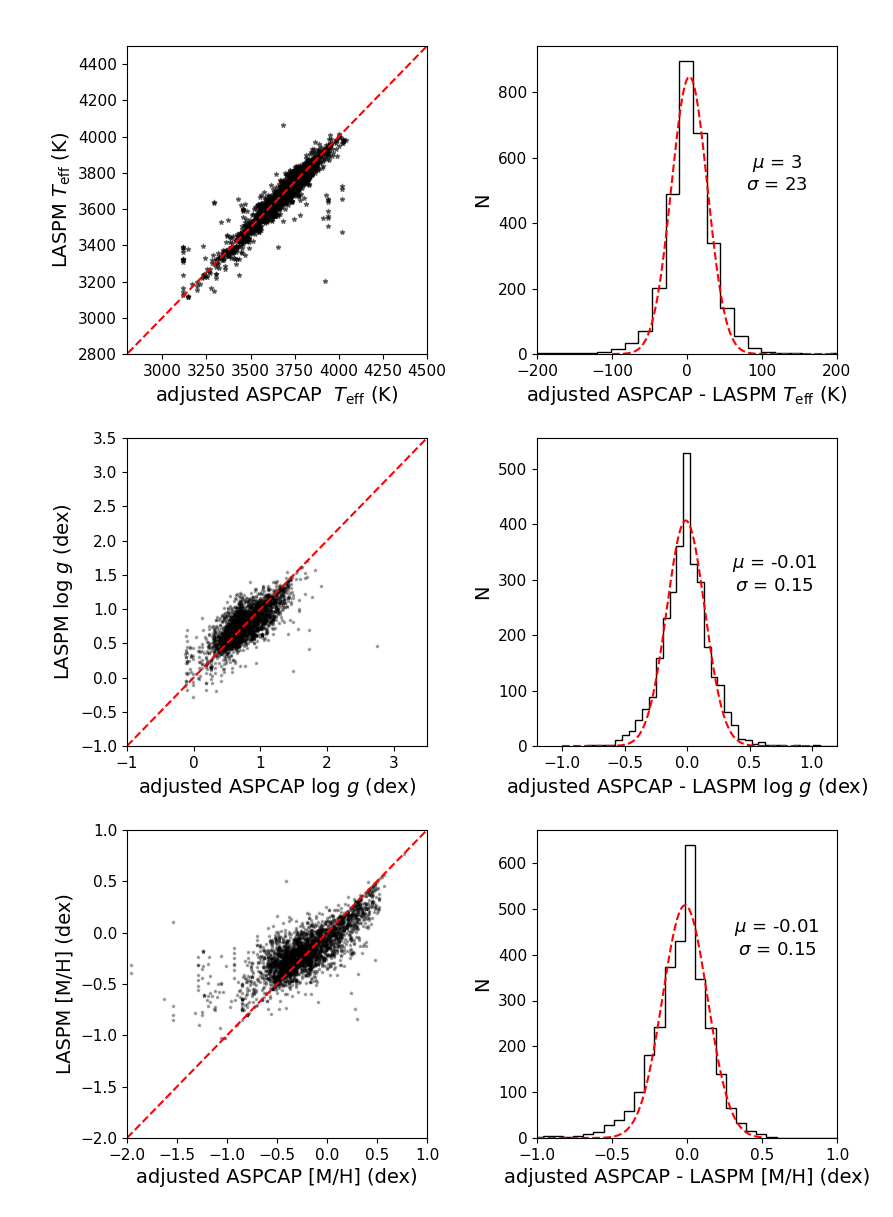}
	\caption{Comparison of the stellar parameters derived from LASPM to the adjusted parameters of ASPCAP for M giants, where adjusted\_\teff \ = teff - 156 (K), adjusted\_\logg \ = \logg \ -0.16 (dex), and adjusted\_[M/H] = [M/H] + 0.17 (dex). The left panel shows one-to-one comparison diagrams and the right panel shows the histograms of differences.}
  \label{fig:compare_APOGEE_giant}
\end{figure*}

Figure \ref{fig:compare_APOGEE_dwarf} shows the comparison between the parameters of LASPM and the adjusted parameters of ASPCAP for M dwarfs. Likewise,  Figure \ref{fig:compare_APOGEE_giant} presents the comparison for M giants. The parameters of LASPM matched the ASPCAP parameters fairly well, with a small scatter of 54 K (23 K) in \teff, 0.06 dex (0.15 dex) in \logg, and 0.12 dex (0.15 dex) in [M/H]  for dwarfs (giants). The LASPM \teff \ of the giant is in good agreement with that of ASPCAP, since the stellar labels for all giant stars are either directly or indirectly based on APOGEE data. The \logg \ values for dwarf stars show a strong consistency due to the concentrated distribution of \logg \ among these dwarfs (4.5--5.0 dex).

\subsection{Comparison with \citet{2024MNRAS.52711866Q}}
\citet{2024MNRAS.52711866Q}, hereafter called Qiu24,  calibrated the metallicity of M dwarfs using wide binary systems consisting of an F, G, or K dwarf paired with an M dwarf. A SLAM model was trained using training data that included metallicity labels from F/G/K companions, with values ranging from $-$1 to 0.5 dex.  Subsequently, the trained SLAM model was applied to determine the metallicity for approximately 630,000 M dwarfs based on the low-resolution spectra from LAMOST DR9.  Compared to APOGEE and other studies that also employed F/G/K+M wide binaries for calibration, the measurements by Qiu24 showed a scatter of $\sim$ 0.18--2.0 dex \citep{2024MNRAS.52711866Q}.

\begin{figure}[ht!]
	\centering
	\includegraphics[width=0.8\linewidth]{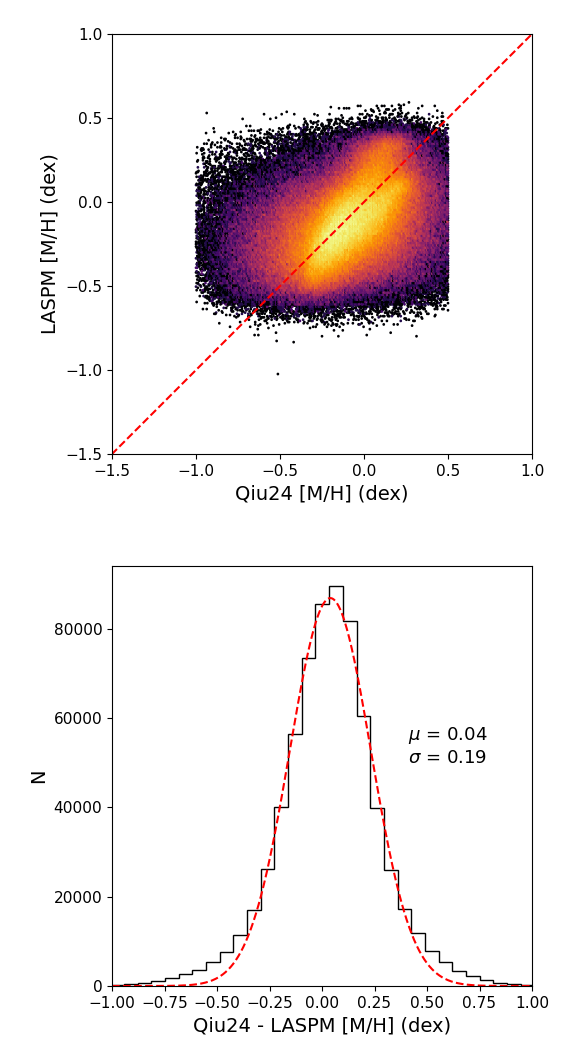}
	\caption{Comparison of the metallicity derived from LASPM to the metallicity of Qiu24. The upper panel shows density plots of direct comparison diagrams and the lower panel shows the histograms of differences.}
  \label{fig:Qiu24_laspm}
\end{figure}

We compared the [M/H]  obtained from LASPM with the metallicity provided by Qiu24, as illustrated in Figure \ref{fig:Qiu24_laspm}.  The metallicities of LASPM were consistent with those calibrated using wide binary systems, exhibiting a bias of 0.04 dex and a scatter of 0.19 dex. As discussed in Section \ref{sec:coverage}, the abundance coverage of our sample leads to an overestimation of metallicity in metal-poor stars with [M/H] $<$ $-$0.6, as illustrated in Figure \ref{fig:Qiu24_laspm}. Ongoing efforts are required to broaden the abundance coverage of our samples.

\clearpage

\section{SUMMARY} \label{sec: summary}

In this work, we present an empirical sample  of 5105 M-type star spectra with  homogeneous atmospheric parameter labels  through stellar label transfer and sample cleaning. That empirical sample is available online in FITS format and in the China-VO Paper Data Repository via\dataset[DOI: 10.12149/101488]{https://doi.org/10.12149/101488}. We chose 1365 spectra with high S/N  from that empirical sample to serve as a reference dataset for LASPM. This reference dataset is additionally available for download via\dataset[DOI: 10.12149/101488]{https://doi.org/10.12149/101488}. The catalog of stellar atmospheric parameters for M-type stars from LAMOST DR11 can be accessed online via\dataset[DOI: 10.12149/101488]{https://doi.org/10.12149/101488} and is also available on the official data release website of LAMOST\footref{fn:dr10-release}.  The 1365 labelled spectra  can serve as an empirical library for the determination of stellar parameters,  and the empirical sample of 5105 M-type stars  offers data annotations for data-driven machine learning models.  The main work of this paper is summarized as follows.

\begin{enumerate}
\item We selected primary samples for M dwarfs and M giants by setting specific distance and magnitude criteria for each to pinpoint the typical stars in the color-magnitude diagram.  Our goal is to guarantee the reliability of parameters for typical stars while excluding rare types, such as subdwarfs, metal-poor stars, and young stellar objects, among others.  We aligned stellar parameters from various catalogs (SDSS DR17 ASPCAP, Qiu23, Ding22, and LASPM) with the StarHorse catalog to correct systematic discrepancies. We used DBSCAN to remove the unreliable samples in each subgrid of parameters to complete the sample cleaning. Further validation was done using a 5-layer neural network, which confirmed the reliability of the stellar labels for the remaining 5105 stars after outlier removal, showing minimal deviation between predicted and actual values.

\item We conducted an internal cross-validation to assess the reliability of stellar labels of the 5105 stars by treating each spectrum as an unknown target and deriving its parameters from the remaining samples. The results showed a scatter of 11 K in \teff, 0.05 dex in \logg, and 0.05 dex in [M/H], respectively, indicating a high precision of the stellar labels. A curated collection of 5105 M-type spectra with well-calibrated fluxes and rest-framed wavelengths,  each with accurately determined stellar labels from multiple sources, has been compiled into an accessible online fits file. 

\item  We chose 1365  spectra  with high S/N  from the empirical sample of 5105 M-type stars to serve as a reference dataset for LASPM, producing an almost seamless Kiel distribution diagram for LAMOST DR11 data. The updated LASPM shows improved precision compared to its predecessor, when S/N $\geq$ 10, with improvements from 118 to 67 K in \teff,  0.2 to 0.07 dex in \logg, and 0.29 to 0.14 dex in [M/H]. We conducted a comparison of LASPM's parameters with the StarHorse catalog across 392,509 stars. The comparison revealed that the parameters from LASPM matched well with those of StarHorse, exhibiting only slight biases. This suggests that LASPM does not introduce global systematic errors in the determination of stellar parameters.  The comparison with APOGEE also showed good agreement, confirming the reliability of LASPM in stellar parameter estimation. The comparison with Qiu24 indicated that the metallicities derived from LASPM were in agreement with those calibrated using wide binary systems. Compared to \teff \ and \logg, the metallicity of M-type stars continues to be the most difficult parameter to determine.

\end{enumerate}
By assembling the spectral library from observed spectra instead of theoretical ones, we to some extent circumvented the difficulties that current spectral synthesis codes face in accurately depicting the complex spectra of cool stars. This empirical sample of M-type stars is very important for many research topics concerning cool stars.

\begin{acknowledgements}


This work is supported by the China Manned Space Project (Nos. CMS-CSST-2021-A08, CMS-CSST-2021-A10, and CMS-CSST-2021-B05), the National Science Foundation of China (grant Nos. 12273078 and 12273057). The authors thank Qiu Dan, Mao-Sheng Xiang, Hong-Liang Yan for helpful discussions. Guoshoujing Telescope (the Large Sky Area Multi-Object Fiber Spectroscopic Telescope LAMOST) is a National Major Scientific Project built by the Chinese Academy of Sciences. Funding for the project has been provided by the National Development and Reform Commission. LAMOST is operated and managed by the National Astronomical Observatories, Chinese Academy of Sciences. This research makes use of data from the European Space Agency (ESA) mission Gaia, processed by the Gaia Data Processing and Analysis Consortium.

\end{acknowledgements}

\end{document}